\documentclass[fleqn,usenatbib]{mnras}

\usepackage{newtxtext,newtxmath}
\usepackage[T1]{fontenc}
\usepackage{ae,aecompl}
\usepackage{verbatim}
\usepackage[dvipsnames]{xcolor}

\usepackage{graphicx}	% Including figure files
\usepackage{amsmath}	% Advanced maths commands
\usepackage{bm}
\usepackage{xcolor}
\usepackage{verbatim}
\usepackage{siunitx}
\usepackage{scrextend} % for the itemize of the dust models
\newcommand{\bmode}{$B$-mode }

\newcommand{\mrm}[1]{\mathrm{#1}}
\title[HWP and beam systematics]{Probing frequency-dependent half-wave plate systematics for CMB experiments with full-sky beam convolution simulations}% in the time-domain}

\author[A. J. Duivenvoorden et al.]{Adriaan J. Duivenvoorden,$^{1}$\thanks{E-mail: adriaand@princeton.edu (AJD)}
Alexandre E. Adler,$^{2}$
Matteo Billi,$^{3,4,5}$\newauthor 
Nadia Dachlythra,$^{2}$
and Jon E. Gudmundsson$^{2}$
\\
$^{1}$Joseph Henry Laboratories of Physics, Jadwin Hall, Princeton University, Princeton, NJ, USA 08544\\
$^{2}$The Oskar Klein Centre, Department of Physics, Stockholm University, AlbaNova, SE-10691 Stockholm, Sweden\\
$^{3}$Dipartimento di Fisica e Astronomia, Alma Mater Studiorum Universit\`a di Bologna, Via
Gobetti 93/2, I-40129 Bologna, Italy\\
$^{4}$INAF-OAS Bologna, Osservatorio di Astrofisica e Scienza dello Spazio di Bologna, Istituto
Nazionale di Astrofisica,\\
via Gobetti 101, I-40129 Bologna, Italy\\
$^{5}$Istituto Nazionale di Fisica Nucleare (INFN), Sezione di Bologna, viale Berti Pichat 6/2, 40127, Bologna, Italy
}

\date{Accepted XXX. Received YYY; in original form ZZZ}

\pubyear{2020}

\begin{document}
\label{firstpage}
\pagerange{\pageref{firstpage}--\pageref{lastpage}}
\maketitle

\begin{abstract}
We study systematic effects from half-wave plates (HWPs) for cosmic microwave background (CMB) experiments using full-sky time-domain beam convolution simulations. Using an optical model for a fiducial spaceborne two-lens refractor telescope, we investigate how different HWP configurations optimized for dichroic detectors centred at \si{95} and \SI{150}{\giga\hertz} impact the reconstruction of primordial $B$-mode polarization. We pay particular attention to possible biases arising from the interaction of frequency dependent HWP non-idealities with polarized Galactic dust emission and the interaction between the HWP and the instrumental beam. To produce these simulations, we have extended the capabilities of the publicly available {\texttt{beamconv}} code. To our knowledge, we produce the first time-domain simulations that include both HWP non-idealities and realistic full-sky beam convolution. Our analysis shows how certain achromatic HWP configurations produce significant 
systematic polarization angle offsets that vary for sky components with different frequency dependence.   
Our analysis also demonstrates that once we account for interactions with HWPs, realistic beam models with non-negligible cross-polarization and sidelobes will cause significant $B$-mode residuals that will have to be extensively modelled in some cases.

\end{abstract}


%%%%%%%%%%%%%%%%%%%%%%%%%%%%%%%%%%%%%%%%%%%%%%%%%%

%%%%%%%%%%%%%%%%% BODY OF PAPER %%%%%%%%%%%%%%%%%%

%\listoftodos
\section{Introduction}
The measured temperature anisotropies of the cosmic microwave background (CMB) provide a large part of the empirical basis for $\Lambda$CDM, the current standard model of cosmology \citep{Boomerang2006, WMAP2013, Planck2018-6}. Additional cosmological information from the CMB will mainly come from accurate characterisation of the polarized component of the anisotropies. Although many cosmological constraints will benefit from polarization measurements \citep{galli_2014}, the most notable advance is perhaps seen in the search for primordial gravitational waves, which might have a distinctive signature in the \bmode component of the CMB polarization~\citep{kamionkowski_1997,Zaldarriaga:1998ar}.  

Experiments have to minimize spurious polarization in order to measure the weak CMB polarization. An attractive approach is the use of a half-wave plate (HWP): a birefringent optical element that shifts the polarization angle of linearly polarized light that passes through. The shift depends on the orientation of the plate, which allows modulation of the polarized sky signal by rotation of the HWP. An ideal rotating HWP only modulates the linearly polarized sky signal and therefore allows one to cleanly separate this desired signal from unpolarized sky signal. Unfortunately, non-ideal HWPs impede perfectly controlled modulation and indirectly cause spurious polarized signal of their own. The merit of a HWP has to be carefully weighed against the downsides. 

 Multiple polarimetric experiments have employed HWPs. Examples include MAXIPOL \citep{Johnson2007}; POLARBEAR \citep{POLARBEAR2010, Hill2016}; ABS \citep{Kusaka2013}; SPIDER \citep{Rahlin2014}; PILOT \citep{Misawa2014}; BLAST \citep{Galitzki2016}; and EBEX \citep{EBEXOptics2018}. In addition, several upcoming \bmode experiments are planning to use HWPs; see e.g.\ the Simons Observatory small-aperture telescopes \citep{Galitzki2018} and the proposed \emph{LiteBIRD} satellite \citep{LiteBIRD2018, Sugai2020}.
Consequently, there exists a rich body of literature describing the optical impact of HWPs, including descriptions of various HWP non-idealities \citep{Bryan2010a, Kusaka2013, Pisano2014, CMBS4-technology} and mitigation strategies \citep{Bao2011, Matsumura2014, Bao:2015eaa, Verges:2020xug}.

In order to separate astrophysical foregrounds from the CMB signal, experiments observe in several frequency bands. For example, the proposed \emph{LiteBIRD} satellite effort currently proposes to deploy 15 frequency bands in three telescope modules spanning 34--\SI{448}{\giga\hertz} \citep{LiteBIRD2018, Sugai2020}. Successful implementation of wide-band polarization modulation is arguably quite technically challenging: the modulation efficiency of simple birefringent crystals is constant over a relatively small frequency range and the plate will cause loss in linear polarization for signals outside that frequency range. In order to efficiently modulate polarization over a wide frequency range, for example to support the use of dichroic or even trichroic bolometers \citep{Suzuki2014}, an achromatic half-wave plate (AHWP) is likely required \citep{Hill2016, LiteBIRDHWP}. AHWPs largely remove the frequency-dependent loss in polarization modulation efficiency, but they can also rotate the polarization angle of linearly polarized light by a frequency-dependent angle. This  angle offset, which can be significant for certain AHWP configurations, is potentially troublesome. When present, an observer needs prior knowledge of the spatial and spectral energy distribution of various astrophysical sources in order to correctly interpret the modulated sky signal. For instance, a sky region dominated by polarized dust requires a different angle correction compared to one dominated by the polarized CMB \citep{Bao2011, abitbol_so_2020}.

In this paper, we investigate how non-idealities from a collection of (A)HWP configurations optimized for dichroic detectors sensitive to both \si{95} and \SI{150}{\giga\hertz} limit our ability to reconstruct primordial $B$-mode polarization. % with a 25\% passband. 
We pay particular attention to the frequency-dependent polarization rotation angle for these different configurations. It has been pointed out, see e.g. \cite{Verges:2020xug}, that such angle offsets will inevitably lead to biased sky maps that require different correcting polarization angles for each sky component. Here, we provide a realistic example of this effect to judge its importance. We also simulate the interaction between the HWP non-idealities and a realistic polarized beam and point out the importance of this potential systematic. 
%and its relation to the varying spectral energy distributions of Galactic dust compared to the CMB.
To produce these simulations, we extend the {\texttt{beamconv}}\footnote{\url{https://github.com/AdriJD/beamconv}} code, first described in \cite{Duivenvoorden2018}. The new code allows us to simulate the effects of non-ideal HWPs on the time-ordered data (TOD) of CMB experiments. To our knowledge, this is the first time that a publicly available code can perform realistic time-domain simulations that include both HWP non-idealities and all-sky beam convolution with asymmetric beams. 

This paper is organized as follows: in Sec.~\ref{sec:MathFramework} we introduce the mathematical framework and the data model used for the simulations. The description of our fiducial instrument, the HWP properties, the proposed scanning strategy and the input sky models are presented in Sec.~\ref{sec:HWPopti}. Results are given in Sec.~\ref{sec:Results}. We discuss the results and formulate our conclusions in Sec.~\ref{sec:Conclusion}.

\section{Mathematical Framework}
\label{sec:MathFramework}
In this section we derive a data model for a typical CMB polarization experiment (see Sec.~\ref{sec:explan_full_M}). The model describes the effects of a non-ideal HWP combined with beam convolution on the time-ordered data. We generalize the model presented in \cite{Bryan2010a} to multi-layer HWPs and arbitrary shaped and non-trivially polarized beams.  
First, however, we briefly discuss the Mueller matrix description of an HWP. See  \cite{Hecht2002o} or \cite{Gil2016pla} for general introductions to the Mueller matrix formalism and e.g.\,  \cite{Bryan2010b, Hileman2013, Moncelsi2014, Salatino2017, Salatino2018} for applications to HWPs for CMB experiments. 

Throughout this section we make use of the Einstein summation convention: pairs of upper and lower indices are implicitly summed over. We use $\theta$ and $\phi$ to denote the polar and azimuthal angles of the standard spherical coordinate system. The metric of the sphere is given by $g_{ij} = \mathrm{diag}(1, \sin^2 \theta )$ in these coordinates.

\begin{figure}
\begin{center}
\includegraphics[width=1.0\linewidth]{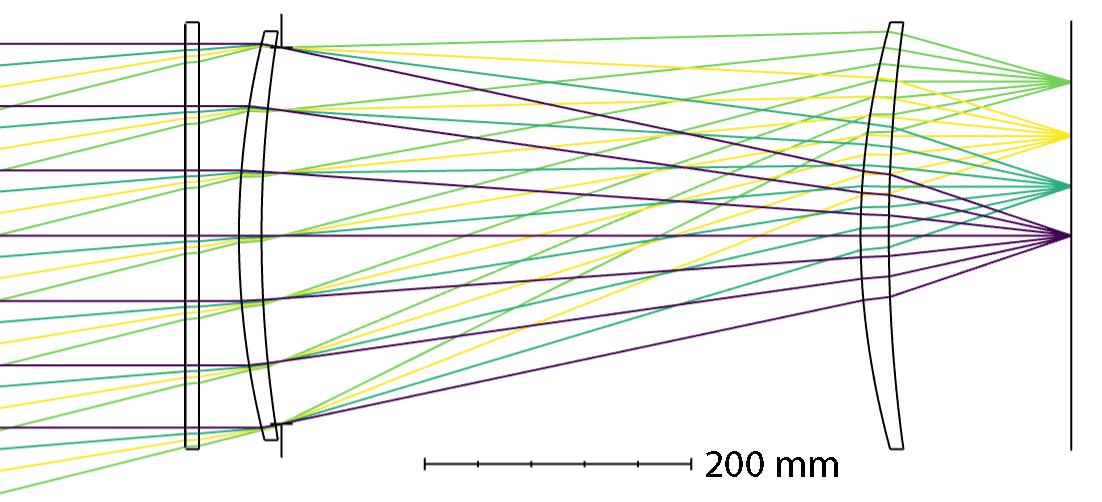}
\caption{\small Sketch of telescope model used for this study. Light coming in from the left interacts with an HWP before hitting the primary lens. Light from the primary lens then gets further focused by the secondary lens before hitting the focal plane (on the right). The edge pixel has a beam centroid of \ang{14} relative to boresight (see ray-bundle emitted from top right corner).}
\label{fig:tel_des}
\end{center}
\end{figure}

\subsection{Half-wave plate Mueller Matrix}

We start by describing the polarized sky signal incident from direction $\hat{\mathbfit{n}}$ and at frequency $\nu$ as a Stokes vector:
\begin{align}
\mathbfit{S}_{\mathrm{sky}}(\hat{\mathbfit{n}}, \nu) = \begin{pmatrix} I \\ Q \\ U \\ V \end{pmatrix} (\hat{\mathbfit{n}}, \nu) \, .
\end{align}
Here, $I$ represents the total intensity of the radiation, while $Q$ and $U$ describe the linearly polarized part of the radiation and $V$ describes the circularly polarized component. Stokes vectors have real elements that obey
\begin{align}
I \geq \sqrt{Q^2 + U^2 + V^2} \, .
\end{align}
The above inequality is saturated for completely polarized light, while the right hand side of the equality goes to zero for unpolarized light. 

Mueller matrices describe the set of linear transformations that transform Stokes vectors to other valid Stokes vectors. Linear optical media such as HWPs are described by Mueller matrices. Multiplying a Stokes vector by such a Mueller matrix describes how the HWP alters the polarization properties of the radiation described by the Stokes vector.

A traditional HWP design involves a single layer of birefringent crystal cut to a thickness such that the phase shift incurred from a particular wavelength at normal incidence is exactly half a period. In the Mueller formalism, a  HWP comprised of a single layer of birefringent material and any number of layers of isotropic dielectric materials can be represented through a matrix characterized by four parameters: 
\begin{equation}
\mathbfss{M}_{\mathrm{HWP}}(\nu) = 
\begin{pmatrix}
T&\rho&0&0\\
\rho&T&0&0\\
0&0&c&-s\\
0&0&s&c
\end{pmatrix} (\nu) \quad \quad \mathrm{(single\ layer)}  \, , 
\label{eq:1lHWP}
\end{equation}
where $T$ can be interpreted as the total transmission, $\rho$ as the difference in transmission between the fast and the slow axes, $c$ as the linear polarisation response and, $s$ as the coupling to circular polarization. The values of these parameters can be directly linked to the Fresnel coefficients for reflection and transmission. For an ideal HWP, we note that $T=1=-c$ and $\rho=s=0$. For a real single-layer HWP these elements are instead variable and dependent on the frequency and the incidence angle of the incoming radiation. 
Fig.~\ref{fig:tel_des} shows how  
the angle of incidence made by light hitting the HWP changes significantly as one moves across the focal plane. For wide field-of-view telescopes, this incidence angle can be as large as \ang{17} \citep{Galitzki2018}. 

\cite{Pancharatnam1955} showed that there exists combinations of layers of birefringent materials that, unlike the single-layer HWPs, can behave in an almost achromatic manner. The resulting achromatic half-wave plates (AHWPs) have a low frequency dependence in polarization modulation efficiency across a broad frequency range. %Such systems introducing a phase delay between the two polarisation modes that stays constant or near constant over a wide range of frequencies. 
This is achieved by introducing a relative rotation angle for one or several of the birefringent layers such that not all of the fast optical axes are aligned. The setup is discussed in detail in \cite{Title1975}. 
A complication of AHWPs is their effective frequency-dependent rotation angle offset. We will come back to this issue in Sec.~\ref{sec:ahwp_phase}.

The Mueller matrix of an AHWP, being composed of more than one birefringent layer, cannot be adequately described by the four parameters in Eq.~\eqref{eq:1lHWP}. 
Instead, the transfer matrix method (TMM) can be used to generate an appropriate Mueller matrix. 
%model representing the response of (infinite) stratified media to incoming plane waves. 
The TMM formalism captures the response of materials that are composed of any collection of dielectric and birefringent media. For the work presented here, we use the publicly available code described in \cite{Hileman2013} to calculate the Mueller matrices of the HWPs that we study.\footnote{\url{ https://github.com/tomessingerhileman/birefringent\_transfer\_matrix}}

\subsection{Data model}
\label{sec:explan_full_M}

We model the TOD of a single detector of a CMB polarimeter as follows:
\begin{align} \label{eq:data_model_gen}
d_t = \int \mathrm{d}\nu \, F(\nu) \int \mathrm{d}\Omega (\hat{\bm{n}}) \, I_{\mathrm{tot}}^{(t)}(\hat{\mathbfit{n}}, \nu) + n_t \, .
\end{align}
The signal incident on the detector $\smash{I_{\mathrm{tot}}^{(t)}}$ depends on the Stokes vector of the sky $\mathbfit{S}_{\mathrm{sky}}$, but is a scalar quantity; the detector is ultimately only sensitive to total intensity. 
The signal is time-varying, the index $t$ runs over the number of recorded time samples.  The frequency passband of the detector and the additive noise are denoted by $F(\nu)$ and $n_t$ respectively.  

To describe how the polarization of the sky couples to the instrument, we express $\smash{I_{\mathrm{tot}}^{(t)}}$ in terms of the trace of the product of two density matrices: one that describes the polarization state of the sky $\mathbfss{W}_{\mathrm{sky}}$ and one time-varying density matrix $\smash{\mathbfss{W}_{\mathrm{instr}}^{(t)}}$ that describes the instrumental response on the sky \citep{hu_2003, Kamionkowski:2015yta, Hivon2017}:
\begin{align}\label{eq:I_det}
I_{\mathrm{tot}}^{(t)}(\hat{\mathbfit{n}}, \nu) = 2 \, \mathrm{tr} \left( \mathbfss{W}_{\mathrm{instr}}^{(t)} \mathbfss{W}_{\mathrm{sky}}\right)  (\hat{\mathbfit{n}}, \nu) \, .
\end{align}
%$\bm{\rho}$ $\rho$ $\mathrm{tr} \left( \mathbfss{W}^{(0)}_{\mathrm{instr}, t} \mathbfss{W}_{\mathrm{sky}}\right)$ $\mathbfss{S}$ $\bm{\mathsf{\rho}}$ $\mathbfss{R}(\psi, \theta, \phi) \hat{\bm{n}}$ 
The density matrices are rank 2 tensor fields defined on the sphere that contain the same polarization state information as the Stokes vectors. In fact, it is possible to express a density matrix $\mathbfss{W}$ in terms of a Stokes vector $S^{\mu} = \{I, Q, U, V\}$ using
\begin{align}\label{eq:stokes2rho}
\mathbfss{W} (\hat{\mathbfit{n}}, \nu) &=  S^{\mu} (\hat{\mathbfit{n}}, \nu) \bm{\mathsf{\sigma}}_{\mu} (\hat{\mathbfit{n}})  \, ,
\end{align}
where $\bm{\mathsf{\sigma}}_{\mu}$ is given by the identity matrix and the (permuted) Pauli matrices defined on the sphere: $\bm{\mathsf{\sigma}}_{\mu} = \{\bm{\mathsf{\sigma}}_0, \bm{\mathsf{\sigma}}_3, \bm{\mathsf{\sigma}}_1, \bm{\mathsf{\sigma}}_2 \}$, see Eqs.~\eqref{eq:pauli_0}-\eqref{eq:pauli_2}. The tensor nature of the polarization state is explicit in the density matrix formulation, it is implicit in the Stokes vector formulation. Using the standard spherical coordinate system, the elements of the sky density matrix are given by
\begin{align}
\big(W_{\mathrm{sky}}\big)_{ij} (\hat{\mathbfit{n}}, \nu) = \frac{1}{2}\begin{pmatrix} I + Q & (U - \mathrm{i} V) \sin \theta \\ (U + \mathrm{i} V) \sin \theta & (I - Q) \sin^2 \theta \end{pmatrix} (\hat{\mathbfit{n}}, \nu) \, .
\end{align}
The time-dependent instrumental density matrix is similarly expressed as
\begin{align}
\begin{split}
&\big(W_{\mathrm{instr}}^{(t)}\big)_{ij}  (\hat{\mathbfit{n}}, \nu) = \\
&\quad \quad \quad \frac{1}{2} \begin{pmatrix} \widetilde{I}^{\,(t)}_{\mathrm{i}} + \widetilde{Q}^{\,(t)}_{\mathrm{i}}  & \big(\widetilde{U}^{\,(t)}_{\mathrm{i}} - \mathrm{i} \widetilde{V}^{\,(t)}_{\mathrm{i}}\big) \sin \theta \\ \big(\widetilde{U}^{\,(t)}_{\mathrm{i}} + \mathrm{i} \widetilde{V}^{\,(t)}_{\mathrm{i}}\big) \sin \theta  &    \big(\widetilde{I}^{\,(t)}_{\mathrm{i}} - \widetilde{Q}^{\,(t)}_{\mathrm{i}}\big) \sin^2 \theta \end{pmatrix}\! (\hat{\mathbfit{n}}, \nu) \, ,
\end{split}
\end{align}
where we have used a tilde to distinguish these Stokes parameters from those of the sky. The $t$ and $\mathrm{i}$ indices denote that the parameters are time dependent and correspond to the instrument (i.e.\ the combination of beam and HWP), respectively.

Both density matrices in Eq.~\eqref{eq:I_det} are defined with respect to the same coordinate basis that is fixed relative to the sky. As a result, the instrumental density matrix $\smash{\mathbfss{W}^{(t)}_{\mathrm{instr}}}$  is time dependent due to the continuous rotation of the instrument with respect to to the sky (another time dependence is due to the HWP rotation, which is kept implicit for now). This time dependence can be factored out by considering the instrumental density matrix in a coordinate system fixed relative to the instrument. Let us denote the density matrix in the instrument frame by $\smash{\mathbfss{W}^{(0)}_{\mathrm{instr}}}$. The two frames are connected by a 3D rotation $\mathbfss{R}_t$ that we define as the rotation that would align the instrument frame to the frame fixed relative to the sky. We can thus perform an active rotation of the $\smash{\mathbfss{W}^{(0)}_{\mathrm{instr}}}$ tensor by $\mathbfss{R}_t$ to get back $\smash{\mathbfss{W}^{(t)}_{\mathrm{instr}}}$: 
%The relation between the instrumental density matrix on the sky $\smash{\mathbfss{W}_{\mathrm{instr}}^{(t)}}$ and the same density matrix in a basis fixed to the instrument $\smash{\mathbfss{W}^{(0)}_{\mathrm{instr}}}$ is given by 
%\begin{align}\label{eq:instr2sky}
%\big(W_{\mathrm{instr}, t}\big)_{ij} (\hat{\mathbfit{n}}) = \Lambda_{i}^{\phantom{a}k} (\mathbfss{R}_t) \Lambda_{j}^{\phantom{a}l} (\mathbfss{R}_t) \big(W^{(0)}_{\mathrm{instr}}\big)_{kl} (\mathbfss{R}_t^{-1} \hat{\mathbfit{n}}) \, .
%\end{align}
\begin{align}\label{eq:instr2sky}
\begin{split}
\big(W_{\mathrm{instr}}^{(0)}\big)_{ij} (\hat{\mathbfit{n}}, \nu) \mapsto& \, \big(W_{\mathrm{instr}}^{(t)}\big)_{ij} (\hat{\mathbfit{n}}, \nu) \\ 
&\,= \Lambda_{i}^{\phantom{a}k} (\mathbfss{R}_t) \Lambda_{j}^{\phantom{a}l} (\mathbfss{R}_t) \big(W^{(0)}_{\mathrm{instr}}\big)_{kl} (\mathbfss{R}_t^{-1} \hat{\mathbfit{n}}, \nu) \, .
\end{split}
\end{align}
The $\bm{\Lambda}$ matrices are matrix representations of the 3D rotation $\mathbfss{R}_t$ 
%that rotates the vector $\hat{\mathbfit{n}}$ from the instrument frame to the frame fixed on the sky 
\citep{challinor_2000}.

The 3D rotation from the instrument frame to the sky frame can be parameterized using 3 time-dependent Euler angles:
\begin{align}\label{eq:rot_in_euler}
\mathbfss{R}_t = \mathbfss{R}(\psi_t, \theta_t, \phi_t) \, .
\end{align}
The $\psi_t$, $\theta_t$, and $\phi_t$ angles can be understood as follows. Imagine a right-handed 3D Cartesian coordinate frame with X, Y, and Z axes centred at the origin of the spherical coordinate system. Let the Z axis point towards the centre of the instrumental response, i.e.\ the beam centre. The 3D rotation is then achieved by a sequence of 3 right-handed rotations: first rotating around the Z axis by the first Euler angle $\psi_t$, then rotating around the Y axis by $\theta_t$ and finally rotating around the Z axis again by $\phi_t$.

Under the rotation $\smash{\mathbfss{R}_t}$ the $\smash{\mathbfss{W}^{(0)}_{\mathrm{instr}}}$ tensor transforms as Eq.~\eqref{eq:instr2sky}. While it is possible to evaluate the transformation directly, we follow \cite{challinor_2000, wandelt_2001} and perform the transformation in the spherical harmonic domain instead. In the harmonic domain, the data model of Eq.~\eqref{eq:data_model_gen} is expressed as follows:
\begingroup
\allowdisplaybreaks
\begin{align}
\begin{split}
d_t &= \int \mathrm{d}\nu F(\nu) \!\sum_{\ell, m, s} \Big\{ b^{\widetilde{I}^{(0)}_{\mathrm{i}}}_{\ell s}\!(\nu, \alpha_t) a^{I}_{\ell m}(\nu) + b^{\widetilde{V}^{(0)}_{\mathrm{i}}}_{\ell s}\!(\nu, \alpha_t) a^{V}_{\ell m}(\nu) \\
&\quad + \frac{1}{2} \Big[ {}_{-2}b^{\widetilde{P}^{(0)}_{\mathrm{i}}}_{\ell s}\!(\nu, \alpha_t) {}_{2}a^{P}_{\ell m} (\nu) + {}_{2}b^{\widetilde{P}^{(0)}_{\mathrm{i}}}_{\ell s}\!(\nu, \alpha_t) {}_{-2}a^{P}_{\ell m} (\nu) \Big] \Big\} \\
& \quad \times  \sqrt{\frac{4 \pi}{2 \ell + 1}} \mathrm{e}^{- \mathrm{i} s \psi_t} {}_s Y_{\ell m} (\theta_t, \phi_t) + n_t\, ,
\end{split}
\label{eq:new_data_model}
\end{align}
\endgroup
where the ${}_sY_{\ell m}$ function is a spin-weighted spherical harmonic and the $\psi_t$, $\theta_t$, and $\phi_t$ Euler angles describe the instrumental pointing. The different $b$ coefficients are spin-weighted spherical harmonic (SWSH) coefficients that describe $\smash{\mathbfss{W}_{\mathrm{instr}}^{(0)}}$, while the different $a$ SWSH coefficients correspond to  $\smash{\mathbfss{W}_{\mathrm{sky}}}$. The sum over $\ell$ runs from $0$ to the harmonic band-limit of the beams: $\ell_{\mathrm{max}}$, while the sums over $m$ and $s$ run from $-\ell$ to $\ell$. It should be noted that the sum over $s$ can be truncated drastically for an approximately symmetric instrumental response. For perfectly symmetric beams only $s=0$ and $s=\pm2$ are needed for the $\smash{\widetilde{I}^{\,(0)}_{\mathrm{i}}}$, $\smash{\widetilde{V}^{(0)}_{\mathrm{i}}}$; and $\smash{\widetilde{P}^{(0)}_{\mathrm{i}}}$ coefficients, respectively \citep{challinor_2000, Hivon2017}.  Exact definitions of the SWSH coefficients are given below and a full derivation is provided in Appendix~\ref{sec:appendixmath}. The expression matches that of a general CMB polarimeter derived in \cite{challinor_2000}, but is generalized to have an explicit dependency on frequency and the HWP rotation angle~$\alpha_t$.   

\begin{figure}
    \centering
    \includegraphics[width=0.48\textwidth]{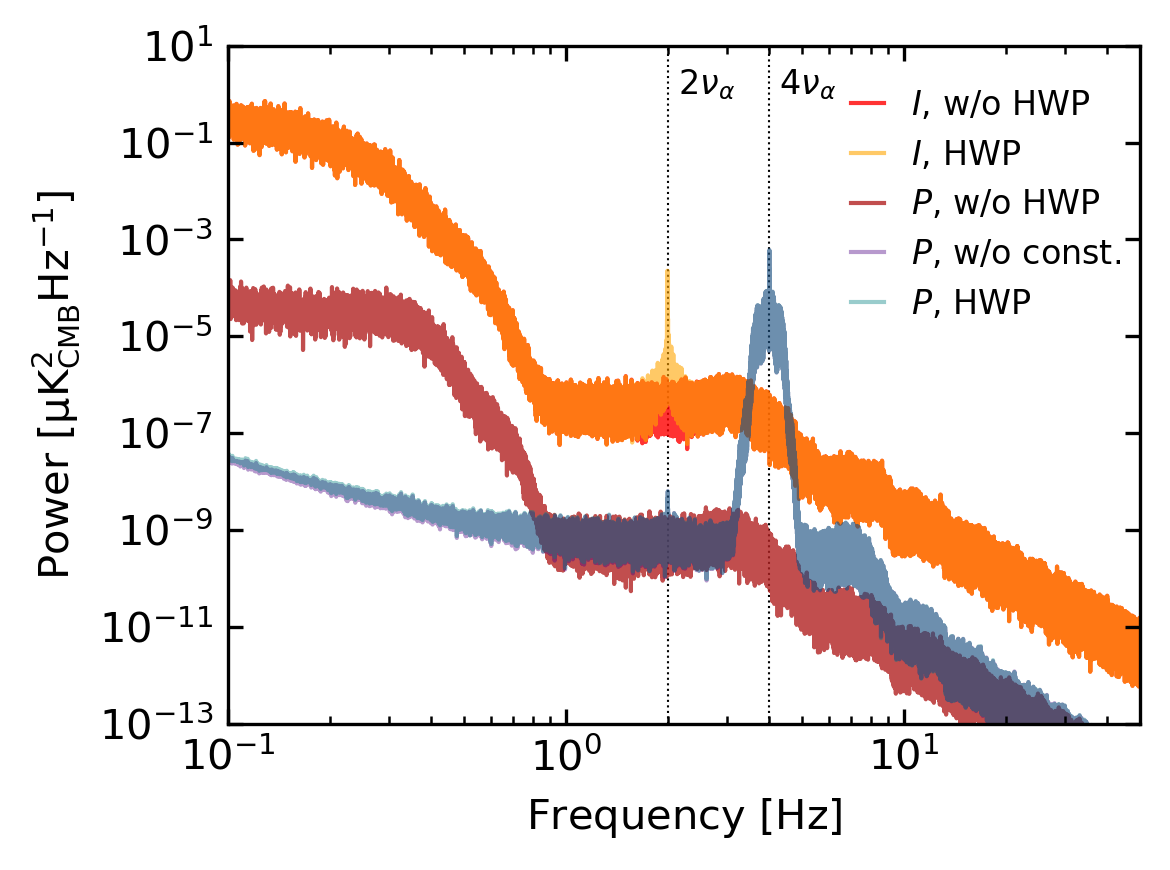}
    \caption{Power spectral densities (PSDs) corresponding to a typical two-hour segment of noiseless time-ordered data for a single detector. The curves labelled \textsf{\textit{I}} (\textsf{\textit{P}}) correspond to scans over an $I$-only (($Q$, $U$)-only) simulated CMB sky. The curves labelled \textsf{HWP} include HWP modulation using the three-layer BR3 HWP configuration (to be discussed in Sec.~\ref{sec:HWPopti})  spinning at a frequency $\nu_{\alpha}$ of \SI{1}{\hertz}. The curve labelled \textsf{\textit{P}, w/o const.} (Overlapping with \textsf{\textit{P}, HWP} but slightly different below $\sim\SI{2}{\hertz}$) incorporates the same HWP modulation, but does not include the HWP systematic that is constant with HWP angle $\alpha$, see Eq.~\eqref{eq:blm_p}. The curves labelled \textsf{w/o HWP} do not include HWP modulation. The simulated data are recorded at a monochromatic frequency of \SI{90}{\giga \hertz} using a Gaussian beam with a FWHM of \SI{32.2}{\arcminute}. Each curve is the average of ten PSDs corresponding to successive two-hour scans. The scan strategy is described in Sec.~\ref{sec:scanstrategy}.}
    \label{fig:psd}
\end{figure}

The $b$ harmonic coefficients that describe the instrument in Eq.~\eqref{eq:new_data_model} are given by combinations of the Stokes parameters of the beam, denoted with the subscript $\mathrm{b}$, and the elements of the HWP Mueller matrix. For the sake of brevity we use a complex representation of the Stokes parameters to describe the linearly polarized beam:
\begin{align}\label{eq:p_beam}
\widetilde{P}^{(0)}_{\mathrm{b}} = \widetilde{Q}^{(0)}_{\mathrm{b}} + \mathrm{i} \widetilde{U}^{(0)}_{\mathrm{b}} \, .
\end{align}
Additionally, we replace the standard HWP Mueller matrix with a complex representation $\mathbfss{C}$ that is indexed by $\{I, P, P^*, V \}$. The two matrices are  related by the following unitary transformation:
\begin{align}
%\bm{\mathcal{M}} 
\mathbfss{C} = \mathbfss{T}
\mathbfss{M}_{\mathrm{HWP}}  \mathbfss{T}^{\dagger} \, ,
\end{align}
where $\mathbfss{M}_{\mathrm{HWP}}$ is the unrotated Mueller matrix and $\mathbfss{T}$ is given by
\begin{align}
\mathbfss{T} = \begin{pmatrix}
1 & 0 & 0 & 0 \\ 
0 & \frac{1}{\sqrt{2}} & \frac{\mathrm{i}}{\sqrt{2}} & 0 \\
0 & \frac{1}{\sqrt{2}} & \frac{-\mathrm{i}}{\sqrt{2}} & 0 \\
0 & 0 & 0 & 1
\end{pmatrix} \, .
\end{align}
The complex representation allows us to cleanly separate terms with different dependence on the HWP rotation angle $\alpha_t$. The harmonic coefficients that describe the instrumental response in Eq.~\eqref{eq:new_data_model} are then given by
\begingroup
\allowdisplaybreaks
\begin{align}
\begin{split}
{}_{\phantom{2}}b^{\widetilde{I}^{(0)}_{\mathrm{i}}}_{\ell s}\!(\nu, \alpha) &=  \int_{S^2}  \mathrm{d}\Omega(\hat{\mathbfit{n}}) \Big[\widetilde{I}^{\,(0)}_{\mathrm{b}}(\hat{\mathbfit{n}}, \nu)  C_{II}(\nu) \\
&\qquad + \widetilde{V}^{(0)}_{\mathrm{b}}  (\hat{\mathbfit{n}}, \nu)  C_{VI}(\nu) \\
&\qquad  + \sqrt{2} \mathrm{Re} \Big( \widetilde{P}^{(0)}_{\mathrm{b}} (\hat{\mathbfit{n}}, \nu)  C_{P^* I}(\nu)\mathrm{e}^{-2i\alpha} \Big) \Big] Y_{\ell s}^* (\hat{\mathbfit{n}}) \label{eq:blm_i} \, ,
\end{split} \\
\begin{split}
{}_{2}b^{\widetilde{P}^{(0)}_{\mathrm{i}}}_{\ell s}(\nu, \alpha) &= \int_{S^2} \mathrm{d}\Omega(\hat{\mathbfit{n}}) \Big[ \widetilde{I}^{\,(0)}_{\mathrm{b}} (\hat{\mathbfit{n}}, \nu)  C_{IP}(\nu) \sqrt{2} \, \mathrm{e}^{-2\mathrm{i}\alpha} \\
&\qquad +  \widetilde{V}^{(0)}_{\mathrm{b}} (\hat{\mathbfit{n}}, \nu)  C_{VP}(\nu) \sqrt{2}  \mathrm{e}^{-2\mathrm{i}\alpha}  \\
& \qquad  +  \widetilde{P}^{(0)}_{\mathrm{b}} (\hat{\mathbfit{n}}, \nu)  C_{P^* P} (\nu) \mathrm{e}^{-4\mathrm{i}\alpha} \\
&\qquad + \widetilde{P}^{(0)*}_{\mathrm{b}} (\hat{\mathbfit{n}}, \nu)  C_{P P}(\nu) \Big] {}_{2}Y_{\ell s}^* (\hat{\mathbfit{n}}) \, , \label{eq:blm_p}
\end{split}   \\
\begin{split}
{}_{\phantom{2}}b^{\widetilde{V}^{(0)}_{\mathrm{i}}}_{\ell s} (\nu, \alpha) &=  \int_{S^2}  \mathrm{d}\Omega(\hat{\mathbfit{n}})  \Big[\widetilde{I}^{\,(0)}_{\mathrm{b}} (\hat{\mathbfit{n}}, \nu)  C_{IV}(\nu) \\ 
&\qquad + \widetilde{V}^{(0)}_{\mathrm{b}}  (\hat{\mathbfit{n}}, \nu) C_{VV}(\nu) \\
&\qquad +  \sqrt{2} \mathrm{Re} \Big( \widetilde{P}^{\,(0)}_{\mathrm{b}}  (\hat{\mathbfit{n}}, \nu)  C_{P^* V}(\nu)\mathrm{e}^{-2\mathrm{i}\alpha} \Big) \Big] Y_{\ell s}^* (\hat{\mathbfit{n}}) \, . \label{eq:blm_v}
\end{split}
\end{align}
\endgroup
The elements of the $\mathbfss{C}$ HWP matrix are given in Eq.~\eqref{eq:complex_elements}. 
Note that the $\smash{{}_{-2}b_{\ell s}}$ coefficients can be obtained using the following symmetry relation:
\begin{align}
{}_{-2}b^{\widetilde{P}^{(0)}_{\mathrm{i}}}_{\ell s}(\alpha) = \big[{}_{2}b^{\widetilde{P}^{(0)}_{\mathrm{i}}}_{\ell -s}(\alpha)\big]^* (-1)^{s} \, .
\end{align}
The harmonic coefficients that represent the Stokes parameters of the sky in Eq.~\eqref{eq:new_data_model} are given by
\begingroup
\allowdisplaybreaks
\begin{align}
a^{I}_{\ell m}(\nu) &= \int_{S^2} \mathrm{d}\Omega(\hat{\mathbfit{n}}) I(\hat{\mathbfit{n}}, \nu) Y^*_{\ell m} (\hat{\mathbfit{n}}) \, , \\
{}_{\pm2}a^{P}_{\ell m}(\nu) &= \int_{S^2} \mathrm{d}\Omega(\hat{\mathbfit{n}}) (Q \pm \mathrm{i} U)(\hat{\mathbfit{n}}, \nu) {}_{\pm2}Y^*_{\ell m} (\hat{\mathbfit{n}}) \, , \\
a^{V}_{\ell m}(\nu) &= \int_{S^2} \mathrm{d}\Omega(\hat{\mathbfit{n}}) V(\hat{\mathbfit{n}}, \nu) Y^*_{\ell m} (\hat{\mathbfit{n}}) \, . \label{eq:alm_v}
\end{align}
\endgroup

Fig.~\ref{fig:psd} helps to qualify the rather verbose expressions for the above harmonic coefficients. It illustrates the effect of a non-ideal HWP on the time-ordered data by comparing the corresponding power spectrum densities for two cases: without an HWP and with a non-ideal HWP (see Sec.~\ref{sec:hwp_selection}). 
Recall that ideal HWP modulation will only modulate the $Q$ and $U$ sky signal, which it will do at a modulation  frequency $4 \nu_{\alpha}$, where $\nu_{\alpha}$ is the HWP rotation frequency. It can be seen that the non-ideal HWP introduces an additional spurious $2 \nu_{\alpha}$ modulation of the $I$ sky (second line of Eq.~\eqref{eq:blm_i}), a $2 \nu_{\alpha}$ modulation of the $Q$ and $U$ sky (first and second line of Eq.~\eqref{eq:blm_p}) and a $2 \nu_{\alpha}$ modulation of the $V$ sky (second line of Eq.~\eqref{eq:blm_v}, not shown in the figure). Finally, the non-ideal HWP also introduces a spurious constant $0 \nu_{\alpha}$  modulation of the $Q$ and $U$ sky (fourth line of Eq.~\eqref{eq:blm_p}).  
Note that Fig.~\ref{fig:psd} omits the case of an input $V$ sky. The $\nu_{\alpha}$ dependence of the $V$-input case will be the same, qualitatively, as the Stokes $I$-input case.

The dependence on HWP angle $\alpha$ of the different terms in the data model is relevant because this dependence is used by the subsequent map-making procedure to distinguish between $I$, $Q$, $U$, (and possibly $V$) sky signal. Leakage between the Stokes parameters will occur when the data model used by the map-maker does not capture the full $\alpha$ modulation of the time-ordered data. For the experimental configuration considered in this work, see Sec.~\ref{sec:HWPopti}, we find that the $I\rightarrow (Q, U)$ leakage that is caused by ignoring the $2\nu_{\alpha}$ terms during map-making is subdominant to the $Q \leftrightarrow U$ leakage that is caused by ignoring  non-idealities in the $4\nu_{\alpha}$ term.

It should be noted that in the derivation of Eqs.~\eqref{eq:blm_i}-\eqref{eq:blm_v} in Appendix~\ref{sec:appendixmath} we have assumed that the instrumental Stokes vector, which is related to $\smash{\mathbfss{W}^{(0)}_{\mathrm{instr}}}$ by Eq.~\eqref{eq:stokes2rho}, can be factored into a Stokes vector describing the beam and a Mueller matrix describing the skywards HWP:
\begin{align}
\label{eq:fact_beam_hwp}
\mathbfit{S}^{(0) \mathsf{T}}_{\mrm{instr}}(\hat{\mathbfit{n}}, \nu, \alpha_t, \vartheta _\mrm{inc}) = \mathbfit{S}^{(0) \mathsf{T}}_{\mrm{beam}} (\hat{\mathbfit{n}}, \nu) \, \mathbfss{M}_{\mrm{HWP}} (\nu, \alpha_t, \vartheta _\mrm{inc}) \, .
\end{align}
The Stokes vector describing the beam has an angular dependence that describes the finite resolution of the experiment, but it is constant with time. On the other hand, the Mueller matrix of the HWP depends on the time-varying HWP angle $\alpha_t$ but is assumed to have no angular dependence. Note that the Mueller matrix varies between detectors based on their position on the focal plane (see Fig.~\ref{fig:tel_des}). This dependence on detector incidence angle is captured by the $\vartheta _\mrm{inc}$ parameter. 
The factorization of the beam and HWP response in Eq.~\eqref{eq:fact_beam_hwp} is an approximation. It allows for separate modelling of the HWP and the instrumental beam. Strictly speaking, the factorization is only valid when the radiation in between the HWP and the beam-forming optical elements is described by plane waves propagating along $\hat{\mathbfit{n}}$. The interaction between the near-field beam and the HWP would in reality also be sensitive to the longitudinal component of the electric field in between the elements. On top of that, the near-field beam is different than the far-field beam described by $\smash{\mathbfit{S}^{(0)}_{\mathrm{beam}}}$. 
Accounting for such near-field effects is beyond the scope of current analysis and simulation infrastructures. We expect that our approximation describes the interaction between the HWP and the beam sufficiently well.

The data model described by Eqs.~\eqref{eq:new_data_model}-\eqref{eq:alm_v} is now implemented in the \texttt{beamconv} library. The frequency dependence of the model is handled by approximating the integral over the instrumental frequency band with a small number ($n_{\nu}=7$ for the results in Sec.~\ref{sec:Results}) of monochromatic input skies, beams and HWP Mueller matrices. The memory costs and computational scaling of the algorithm have thus gained a linear scaling with $n_{\nu}$ compared to the algorithm in \cite{Duivenvoorden2018} but are unchanged otherwise. The algorithm allows for efficient time-domain simulations that include all-sky beam convolution with asymmetric beams and non-ideal HWPs.

\section{Simulation setup}
\label{sec:HWPopti}
\begin{figure}
    \centering
    \includegraphics[width=0.48\textwidth]{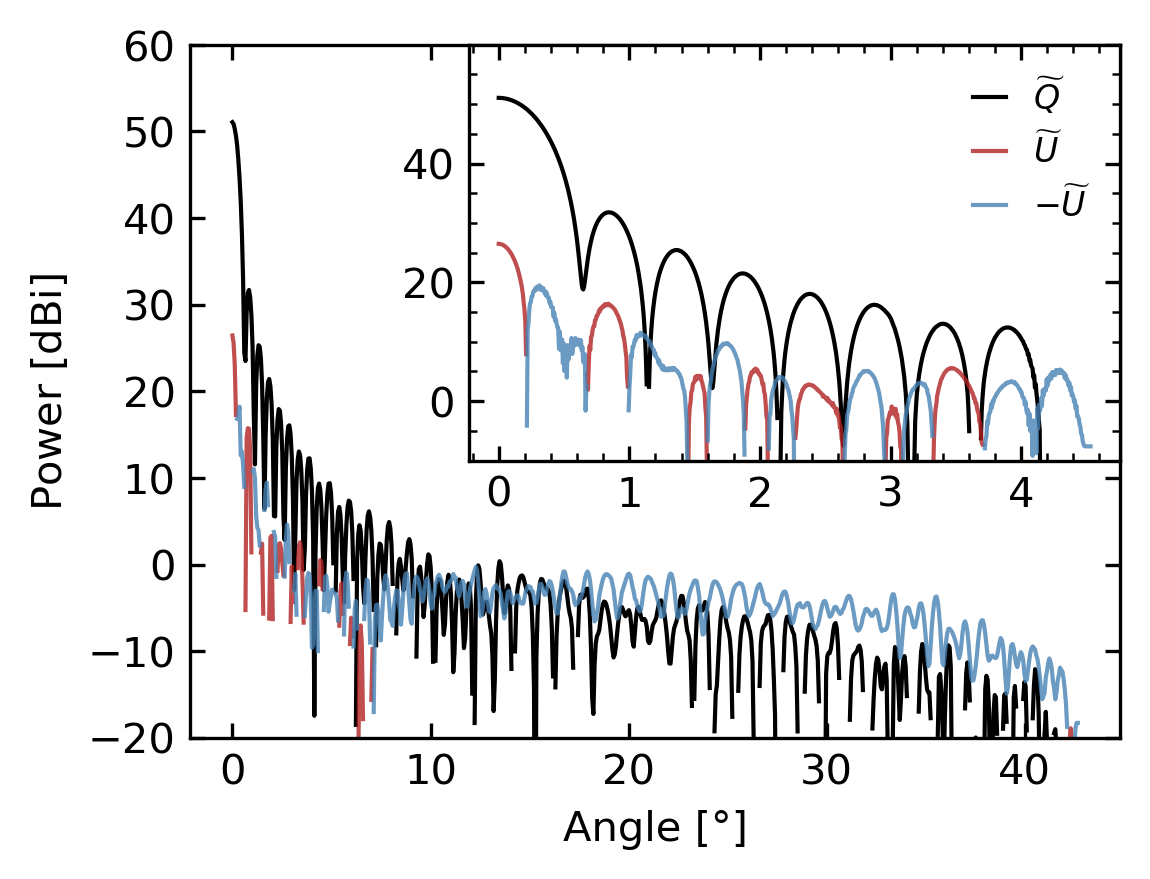}
    \caption{Azimuthally averaged beam profiles (dBi units) for a representative detector of one of the 50 used in this analysis.  
     Shown are the Stokes $\widetilde{Q}$ and $\widetilde{U}$ beam components.
     For this figure, we have defined the Stokes parameters with respect to the Ludwig-3 basis \citep{ludwig_1973}. This basis is approximately Cartesian  around the beam centre and has been aligned with the polarized element of the detector. As a result, the $\pm \widetilde{U}$ profile quantifies the amount of non-aligned (or ``cross-polar'') polarized sensitivity of the beam. It can be seen that $|\widetilde{U}|$ is subdominant close to the centre of the beam (see inset) while having a relatively large contribution at large opening angles.}
    \label{fig:profile}
\end{figure}

We consider a telescope similar to the one described in \citep{Duivenvoorden2018}, but with a HWP in front of the primary lens. Incoming radiation passes through the HWP followed by a pair of lenses before being absorbed by the detectors on the focal plane (see Fig.~\ref{fig:tel_des}). A beam profile for a typical \SI[number-unit-product=\text{-}]{150}{\giga\hertz} detector used in this analysis is shown in Fig.~\ref{fig:profile}. We model 50 dichroic detectors sensitive to two \SI[number-unit-product=\text{-}]{30}{\giga\hertz}-wide frequency windows centred at 95 and \SI{150}{\giga\hertz}. The detectors are evenly distributed on a square grid of a focal plane fed by a \SI[number-unit-product=\text{-}]{30}{\centi\metre} aperture telescope. The field of view of this square grid is only \ang{7} compared to the \ang{28} that can be supported by this telescope; the detectors therefore only cover a fraction of the focal plane. The spectral response of the detectors is assumed to be represented by a top-hat function within each band. In order to test frequency dependent effects, we run simulations at 7 sub-frequencies within a band. These sub-frequencies are 80, 85, 90, 95, 100, 105, and \SI{110}{\giga\hertz} for the \SI[number-unit-product=\text{-}]{95}{\giga\hertz} band and 135, 140, 145, 150, 155, 160, and \SI{165}{\giga\hertz} for the \SI[number-unit-product=\text{-}]{150}{\giga\hertz} band (see hatched regions in Fig.~\ref{fig:MM_elms}). 

\subsection{Simulated scanning}\label{sec:scanstrategy}

Using the updated version of {\tt beamconv}, we simulate one year of satellite scanning for 50 detectors. We use a similar scan strategy as in \cite{Duivenvoorden2018}, which is based on \cite{gorski_2008, Wallis2017}. The satellite spins around its principal axis with a period of 600 seconds. It precesses about the boresight axis with a period of 90 minutes. The two axes are separated by \ang{50}. We set the HWP rotation frequency $\nu_{\alpha}$ to \SI{1}{\hertz} (angular frequency of $\SI{2 \pi}{\radian / s}$) and sample the data at \SI{12.01}{\hertz}. Although the sampling frequency is likely an order of magnitude below that of a real experiment, we find that this rate suffices for our noiseless simulations. The resulting angular coverage is excellent and allows for simultaneous per-pixel recovery of $I$, $Q$, and $U$ over the full sky. Even without a continuously-spinning HWP, the average condition number of the per-pixel $(I, Q, U)$ covariance matrix, which is inverted as part of the solution \citep{Duivenvoorden2018}, is approximately 2.9 for a $N_\mrm{side} = 256$ map. In comparison, the condition number approaches 2.0 (the minimum value) for all pixels when the HWP is spun with a \SI[number-unit-product=\text{-}]{1}{\hertz} rotation frequency.

\subsection{Input maps}

We generate statistically isotropic random Gaussian Stokes $I$, $Q$, and $U$ CMB maps (with a vanishing \bmode component) using the {\tt synfast} utility in {\tt HEALPix}'s \citep{Healpix2005} Python implementation, {\tt healpy}\footnote{\url{http://healpix.sf.net}}\footnote{\url{https://github.com/healpy/healpy}} and the best-fit 2018 \emph{Planck} power spectra \citep{Planck2018-6}. 
To probe how frequency-dependent HWP systematics interact with the different components of the microwave sky, we also simulate polarized Galactic dust using the Python Sky Model ({\tt PySM}) code \citep{PySM}. Other foreground sources, including synchrotron radiation, are subdominant in our \SI{95} and \SI{150}{\giga\hertz} frequency bands. {\tt PySM} provides different templates for dust emission, all based on the high-frequency \emph{Planck} data \citep{Planck2016-10}.\footnote{\url{https://pysm3.readthedocs.io/en/latest/}} 
We use six different \texttt{PySM} dust models: \texttt{d0} to \texttt{d5}. The first four models are directly based on a modified black body distribution. In units of CMB brightness temperature these models all follow the same parametrization:
\begin{align}\label{eq:mbb_dust}
\begin{pmatrix}Q \\ U \end{pmatrix} (\hat{\mathbfit{n}}, \nu) =\begin{pmatrix}A_{Q} \\ A_{U} \end{pmatrix} \! (\hat{\mathbfit{n}})  \times \left(\frac{\nu}{\nu_0}\right)^{\beta(\hat{\mathbfit{n}}) + 1} \frac{\mathrm{e}^{h\nu_0 / k_{\mathrm{B}}T(\hat{\mathbfit{n}}) }-1}{\mathrm{e}^{h\nu / k_{\mathrm{B}}T(\hat{\mathbfit{n}}) }-1} \, ,
\end{align}
There are four parameters: the spectral index $\beta$, the dust temperature $T$ and the $A_{Q/U}$ amplitudes at the reference frequency $\nu_0 = \SI{353}{\giga \hertz}$. A brief description of each model follows, see \cite{PySM} for more details.
\begin{labeling}{\textbf{d1}}
    \item[\texttt{d0}] uses a fixed spectral index ($\beta=1.54$), a fixed temperature ($T=\SI{20}{\kelvin}$) and the \texttt{Commander} dust template from  \cite{Planck2015_diffuse_foreground} for $A_{Q/U}$.
    \item[\texttt{d1}] extends the \texttt{d0} model with spatially varying  spectral index and temperature that are both given by the \texttt{Commander} templates from  \cite{Planck2015_diffuse_foreground}.
    \item[\texttt{d2}] modifies the d1 model with a spectra index that varies randomly on degree scales, following a Gaussian distribution: $\beta \sim \mathcal{N}(\mu=1.59, \sigma^2 = 0.04)$.
    \item[\texttt{d3}] is the same as \texttt{d2} except that  $\beta \sim \mathcal{N}(\mu=1.59, \sigma^2 = 0.09)$.
    \item[\texttt{d4}]  models two dust populations as two modified black bodies with different but spatially constant spectral indices and two different spatially varying temperatures and dust amplitudes \citep{Finkbeiner2015}.
    \item[\texttt{d5}] is a more physically motivated model based on the physical properties of two populations of dust grains (silicate and carbonaceous) \citep{Hensley2015, Hensley:2017ygd}.
\end{labeling}  
The inclusion of these six models in our analysis serves to roughly bracket the current uncertainty in dust modelling. We note that the \texttt{d3} model is designed to match the largest variation in spectral index allowed by the \emph{Planck} data. We study the interplay between the HWP non-idealities and these different foreground models in Sec.~\ref{sec:fgdep}. 

\subsection{Selection of HWPs}\label{sec:hwp_selection}

\begin{figure*}
%\centering
\includegraphics[width=\textwidth]{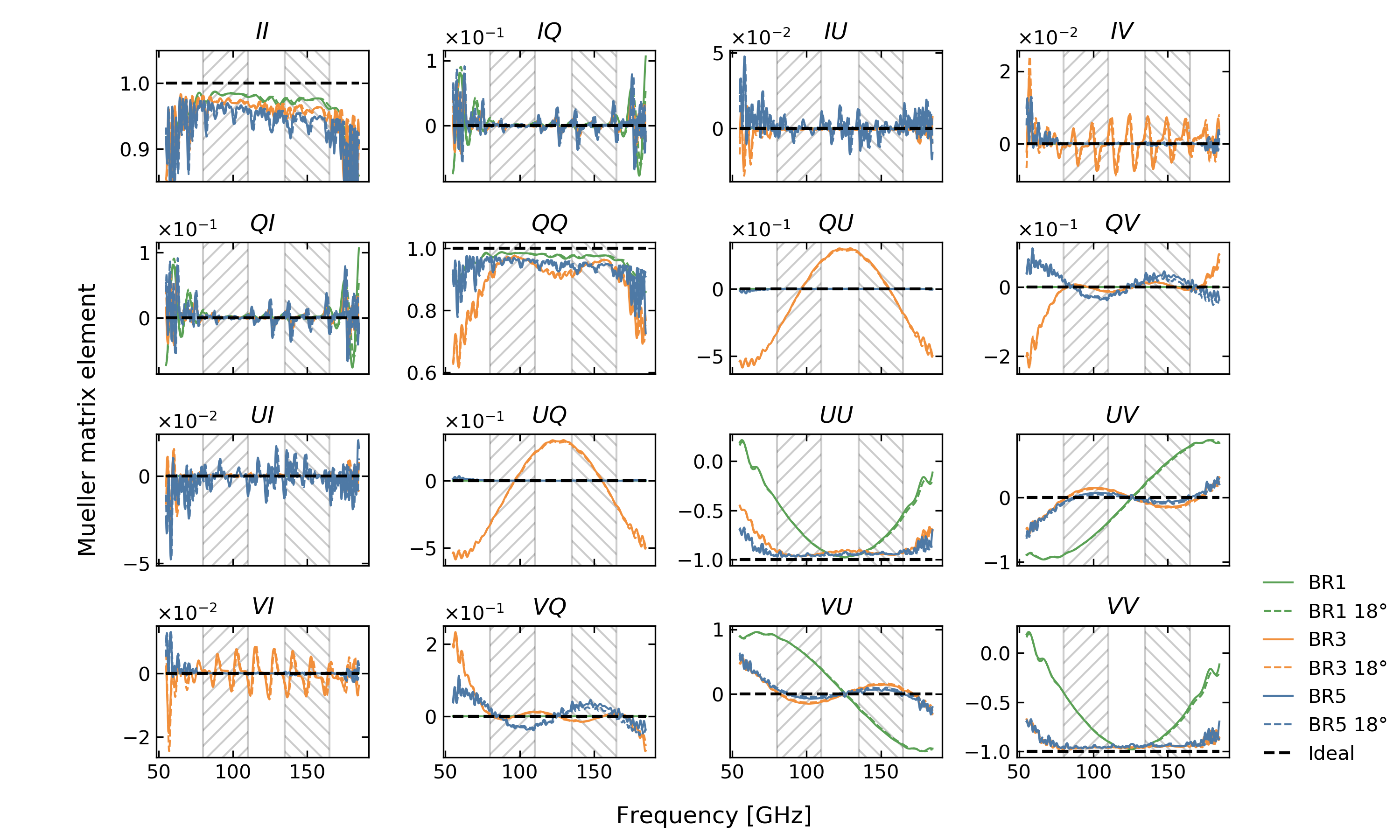}
\caption{\label{fig:MM_elms} HWP Mueller matrix elements as a function of frequency in the normal incidence case (solid lines) and for an incidence angle $\vartheta_{\mathrm{inc}}$ of \ang{18} (dashed lines, virtually indistinguishable from solid lines) simulated using the transfer matrix method. The three HWP configurations described in Table~\ref{table:HWPconfig} are shown. A \ang{31.4} HWP rotation angle offset is applied to the 3-layer BR3 model. The black dashed line represents the ideal HWP ($T=-c=1$, $\rho=s=0$ in Eq.~\eqref{eq:1lHWP}). The grey hatched bands illustrate the two instrumental frequency bands used in this work. 
}
\end{figure*}

\begin{figure*}
%\centering
\includegraphics[width=\textwidth]{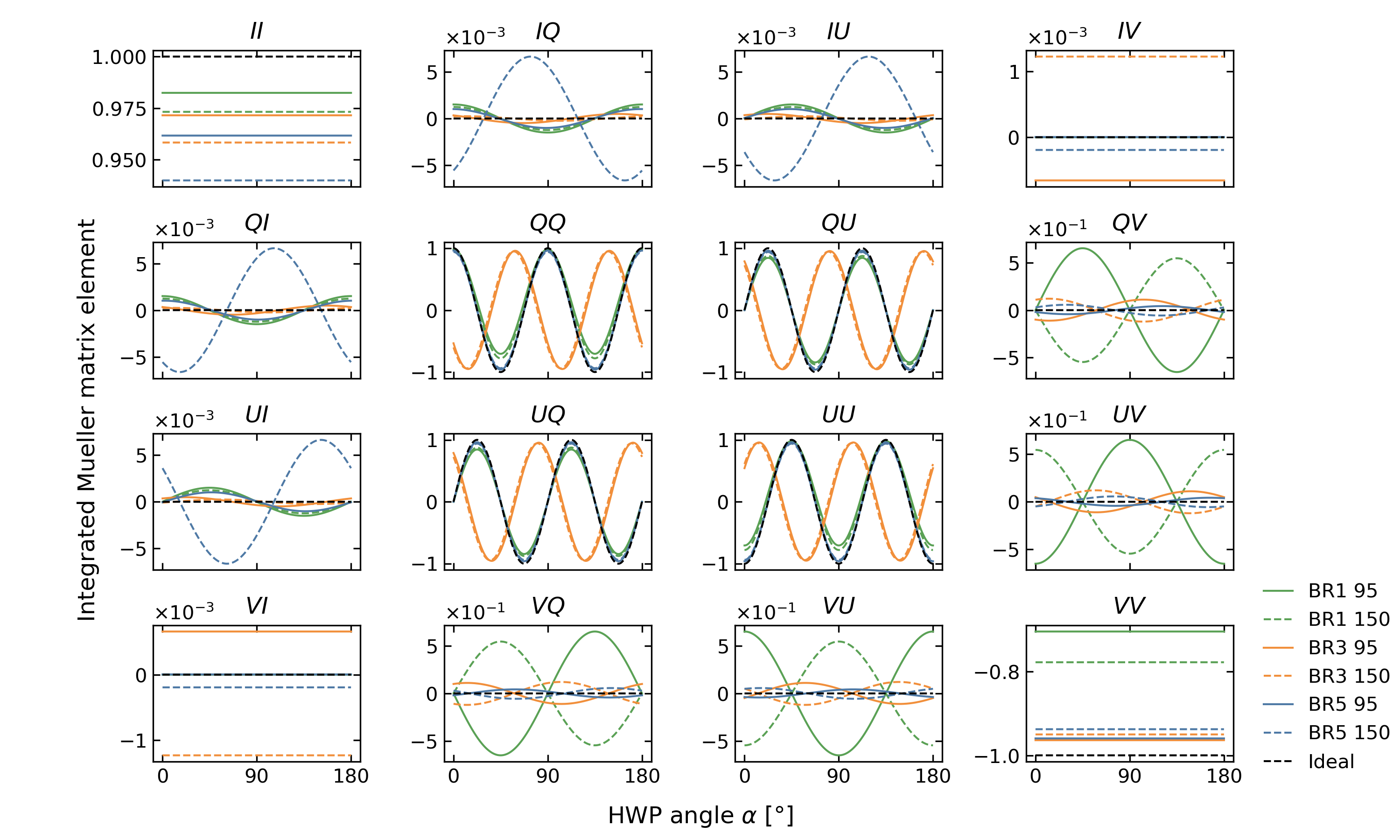}
\caption{\label{fig:MM_elms_alpha} Mueller matrix elements for the three HWP models described in Table~\ref{table:HWPconfig}, integrated over the instrumental frequency bands (\textsf{95}: solid lines, \SIrange{80}{110}{\giga\hertz}; \textsf{150}: dashed lines,  \SIrange{135}{165}{\giga\hertz})  as a function of the HWP rotation angle. The dashed black lines represent the behaviour of the ideal HWP ($T=-c=1$, $\rho=s=0$ in Eq.~\eqref{eq:1lHWP}). It can be seen that the BR3 configuration (orange lines) is out of phase with the other HWP configurations.}
\end{figure*}

\label{sec:choosing_hwp}

\begin{table}
\begin{tabular}{lccc}
\hline
Model & Orientation & Phase \SI{95}{\giga\hertz} & Phase \SI{150}{\giga\hertz} \\
& & CMB/Dust & CMB/Dust
\\\hline \hline
BR1 & \ang{0} & \ang{0}/\ang{0} & \ang{0}/\ang{0}\\ \hline
BR3 & $\{\ang{0}, \ang{54},\ang{0} \}$ &  \ang{30.75} / \ang{31.16}
& \ang{32.51} /  \ang{32.30}
\\\hline
BR5 & $\{\ang{22.9}, \ang{-50}, \ang{0}$ & & \\ & $\ang{50}, \ang{-22.9}\}$& \ang{0}/\ang{0} & \ang{0}/\ang{0}\\\hline
\end{tabular}
\caption{HWP configurations adopted for the analysis presented in this paper. Orientation angles are those of the fast axis of the birefringent layers relative to the plane of incoming vertically polarised radiation. The rotation angle offset is given in each band following Eq.~\eqref{eq:Rhwp}, for CMB and dust weights as defined in Eq.~\eqref{eq:dustweight}.
\label{table:HWPconfig}
}
\end{table}

A wide range of HWP designs have been described and studied in the literature \citep{Bryan2010a,Hill2016} \citep{LiteBIRDHWP,EBEXOptics2018,POLARBEAR2010}. HWP design involves a complex optimization problem where absorptive and reflective losses from materials with high index of refraction need to be balanced against the desire for unity polarization efficiency across a wide band. We choose to study three HWP configurations, which are loosely based on \citep{Bryan2010a} as a model of a one layer HWP, \citep{Hill2016} for the 3-layer HWP, and a 5-layer HWP model taken from \citep{Komatsu2020}. Some key properties of these three HWP configurations, which we denote as BR1, BR3, and BR5, are shown in Table \ref{table:HWPconfig}. 

We adopt a fixed thickness, $d=\SI{3.75}{\milli\meter}$, for the individual sapphire plate layers for all three polarisation modulators. This thickness was found using the traditional formula for half wave plates made of a single layer of birefringent material $d = c/\left[2\nu (n_\mrm{e} - n_\mrm{o})\right]$, where $n_\mrm{o}$ and $n_\mrm{e}$ correspond to the index of refraction for the ordinary and extraordinary axes, respectively. The selected thickness is optimal for $\nu = \SI{126}{\giga\hertz}$, near the average of our two band centres. We adopt an anti-reflection coating similar to the one described in \cite{Coughlin2018} that is optimized for 75--\SI{170}{\giga\hertz}.  We settle on three AR layers with thicknesses $d_\mrm{AR} = 0.5, 0.31, \SI{0.257}{\milli\meter}$ and individual indices $n_\mrm{AR} = (1.268, 1.979, 2.855)$. The above parameters are used as input to the TMM formalism to calculate the Mueller matrices of the HWPs. We produce a unique set of Mueller matrices for each unique HWP incidence angle $\vartheta_{\mathrm{inc}}$.  

Figure \ref{fig:MM_elms} shows the Mueller matrix elements for our three HWP configurations as function of frequency. It can be seen that the additional layers of the BR3 and BR5 HWPs improve the frequency uniformity of the polarization efficiency (see the $UU$ elements) compared to the BR1 case. Describing the efficiency loss for the different  Stokes parameters is a rather complicated task. Although the efficiency loss of Stokes $I$ is easy to understand, as the $II$ elements decrease in value with additional layers, the same is not true for the polarization efficiency.\footnote{The amplitude of incoming linear polarization $\sqrt{Q^2 + U^2}$ will be changed based on the $QQ$, $QU$, $UQ$, $UU$ submatrix. The change in amplitude will be bounded by the singular values of this matrix. Note that the amplitude change will generally be different per pixel and frequency. Furthermore, the input $I$ and $V$ signal will also alter the linear polarization amplitude due to leakage caused by the $QI$, $UI$, $QV$ and $UV$ terms.}
Because of these complications, we do not directly use the HWP Mueller matrix elements to correct our results for the efficiency loss. As will be detailed in Sec.~\ref{sec:Results}, we settle for a more robust and simpler power-spectrum based calibration method. Such an approach will likely also be taken by a real experiment. 
Finally, we note that the Mueller matrix models that we use do not include systematic effects caused by non-ideal manufacturing or material non-uniformity, which are likely to exist at some non-negligible level even in next-generation experiments.

\subsection{Determining the AHWP induced rotation offset}
\label{sec:ahwp_phase}

\begin{figure}
    \centering
    \includegraphics[width=.47\textwidth]{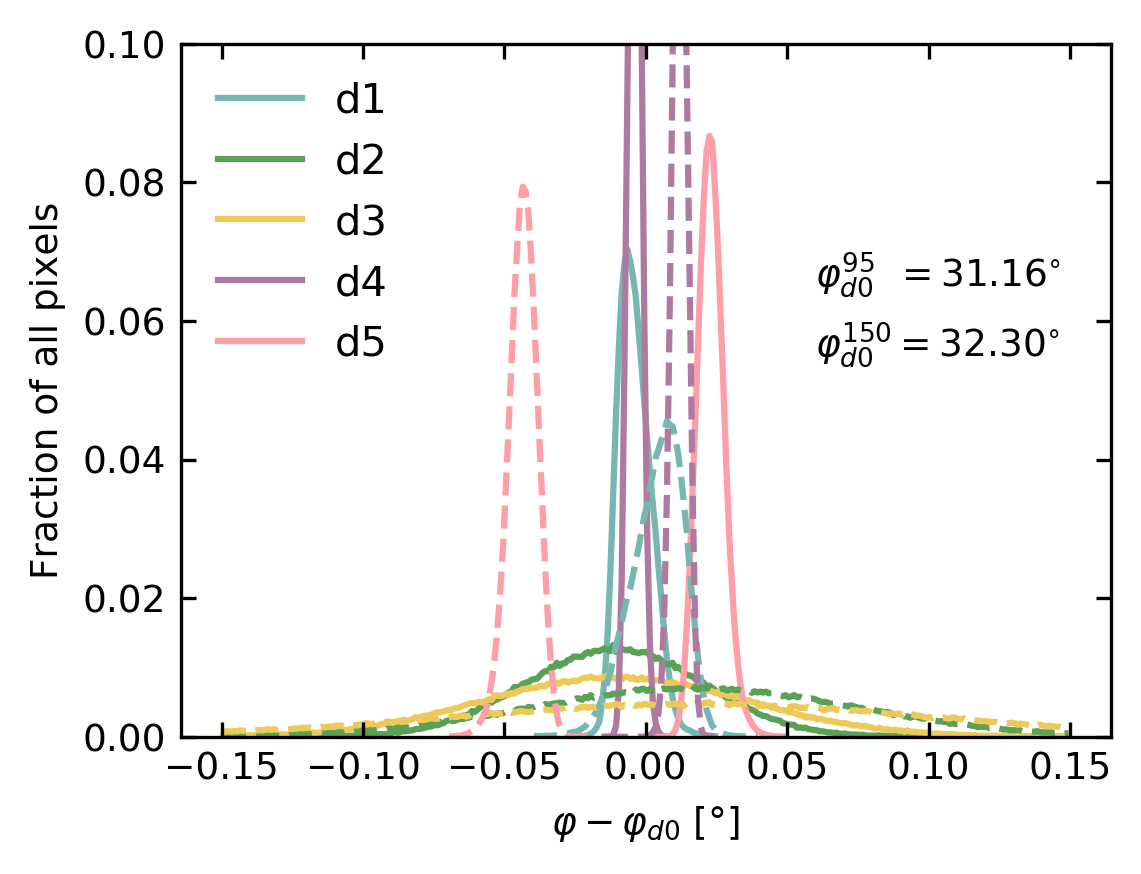}
    \caption{
    Distribution of optimal BR3 HWP rotation angle offset $\varphi$ in the 95 (solid lines) and \SI{150}{\giga\hertz} (dashed lines) bands for the \texttt{PySM} Galactic dust models based on their per-pixel spectral energy distribution at $N_\mrm{side}=512$. The distributions are given for the $40\%$ sky mask used in our analysis. The abscissa is expressed as the difference between the rotation angle offset $\varphi$ and the reference angle $\varphi_{d0}$ corresponding to a modified blackbody with $T=\SI{20}{\kelvin}$ and $\beta=1.54$ as in Table~\ref{table:HWPconfig}. %Bin width is $\ang{0.001}$
    }
    \label{fig:dustanglehist}

\end{figure}

Achromatic HWPs, such as the three- and five-layer configurations discussed in this paper, tend to have higher polarization efficiency over a given frequency range compared to a single-layer HWP. However, they also introduce an undesirable frequency-dependent phase between the in-going and out-going electric field that manifests itself as a frequency-dependent HWP rotation angle offset. Fig.~\ref{fig:MM_elms_alpha} shows our HWP Mueller matrices, integrated over the two frequency bands, as a function of the half-wave plate angle $\alpha$. From the inner two-by-two set of panels it is clear that the 3-layer HWP has a relatively large rotation angle offset. It turns out that the offset angle of the 3-layer model also displays the largest variation with frequency. While the average value of this offset angle can be simply calibrated out, this large variation with frequency poses a difficulty: sky components with different frequency characteristics will require different offset angles after integration over the instrumental frequency band. 

We can determine an optimal rotation angle offset for a specific sky component as the HWP rotation angle, $\alpha_{\mathrm{min}}$, that minimizes the difference between the $QQ$, $QU$, $UQ$, $UU$ submatrices of the Mueller matrices of the HWP and the ideal HWP. The $\alpha_{\mathrm{min}}$ angle is found by minimising
\begin{equation}
   R(\alpha) =   \sum_{i,j \in \{Q, U \}} \left[  \sum_{k=1}^{n_{\nu}}  w(\nu_k) \: M_{\mathrm{HWP},ij}(\nu_k) - D_{ij}(\alpha) \right]^2 \, , \label{eq:Rhwp}
\end{equation}
where $\mathbfss{M}_{\mathrm{HWP}}(\nu_k)$ 
is the same as in Eq.~\eqref{eq:fact_beam_hwp}  with normally incident light and $\mathbfss{D}(\alpha)$ is the Mueller matrix of the ideal HWP rotated by an angle $\alpha$. The $\nu_k$ are a set of sub-frequencies within the band, and $w(\nu_k)$ are weights applied to model the SED. Because we work in units of CMB brightness temperature, we use uniform weighting for the CMB. If we assume that Galactic dust follows a modified blackbody distribution with a fixed temperature and spectral index across the sky, the weights can be derived from Eq.~\eqref{eq:mbb_dust}:
\begin{equation}
        w(\nu_k) = \left(\sum_{i=1}^{n_{\nu}} \frac{\nu_i^{\beta+1}}
        {\mathrm{e}^{h\nu_i  / k_{\mathrm{B}} T} -1 } \right)^{-1} \frac{\nu_k^{\beta+1}}
        {\mathrm{e}^{h\nu_k / k_{\mathrm{B}}T}-1   
        } \, .\label{eq:dustweight}
\end{equation} 
Note however that these assumptions about the dust SED are only valid for the {\tt d0} {\tt PySM} model (with  $T=\SI{20}{\kelvin}$ and $\beta=1.54$). The optimal offset angles for the CMB and the above dust weights are given in Table~\ref{table:HWPconfig}. The 3-layer configuration shows a significantly different optimal offset angle for the CMB versus dust.

The optimal HWP rotation angle correction will vary across the sky for foregrounds models that include spatial SED variations. We can determine an optimal per-pixel correction for a given foreground component by applying Eq.~\eqref{eq:Rhwp} on a pixel-by-pixel basis. In Fig.~\ref{fig:dustanglehist} we compare the distribution of the optimal HWP rotation offset angles for the \texttt{d1}-\texttt{d5} {\tt PySM} dust models to the \texttt{d0} value given by Eq.~\eqref{eq:dustweight}. 
We only show results for BR3 in Fig.~\ref{fig:dustanglehist}. The BR1 and BR5 configurations have a near-constant rotation angle offset over the range of frequencies that we consider and show no appreciable deviation from an isotropic angle offset. Calculating the distributions in Fig.~\ref{fig:dustanglehist} requires knowledge on the per-pixel SED weights $w(\nu_k)$ in Eq.~\eqref{eq:Rhwp}. 
Although we lack a closed-form expression for all of the SEDs of our dust models, we can make use of the \texttt{PySM} predictions at each subfrequency $\nu_k$ to determine the SED weights using 
\begin{equation}
    w(\hat{\mathbfit{n}}, \nu_k) = \Bigg(\sum_{j=1}^{n_{\nu}} \lvert P(\hat{\mathbfit{n}}, \nu_j) \rvert \Bigg)^{-1} \lvert P(\hat{\mathbfit{n}}, \nu_k) \rvert\, ,\label{eq:varphiweights}
\end{equation}
where $\lvert P(\hat{\mathbfit{n}}, \nu_k) \rvert$ is the amplitude of linear polarization at subfrequency $\nu_k$ in direction $\hat{\mathbfit{n}}$.

\section{Analysis Results}
\label{sec:Results}

% Description of the simulation output and the pipeline:
% Sub-frequency maps, Planck mask, PolSPICE, beam deconvolution
To test the capabilities of the updated {\tt beamconv} code, we run a number of simulations that probe the different HWP configurations, sky models and instrumental beams. Each simulation batch is based on seven sub-frequency maps per frequency band that are combined assuming a top-hat passband. Seven sub-frequencies represent the lowest adequate sampling of the frequency variation of the HWP Mueller matrices. The simulated time-ordered data are binned on the sphere using the standard map-making scheme that ignores the instrumental beam and assumes the following data model for each detector:
\begin{align}
\begin{split}
d_{t} = I(\hat{\mathbfit{n}}_t) + Q(\hat{\mathbfit{n}}_t) \cos&\big[2 (\psi_t + \gamma) + 4 (\alpha_t + \varphi)\big] \\
+ U(\hat{\mathbfit{n}}_t) \sin&\big[2 (\psi_t + \gamma) + 4 (\alpha_t + \varphi)\big] + n_t\, .
\end{split}\label{eq:mapmaker}
\end{align}
Here, $\hat{\mathbfit{n}}_t$, $\psi_t$ and $\alpha_t$ describe the instrumental pointing and HWP rotation angle at time-sample $t$ while $\gamma$ and $\varphi$ describe the detector polarization angle and HWP rotation angle offset, respectively. 
The map-maker solves for $I$, $Q$ and $U$ per pixel, uses uniform weighting of the time-ordered data and does not explicitly use detector pair differencing, see e.g.\ \cite{Duivenvoorden2018}.
%insert mapping info here

For every simulated systematic effect, the same simulation is performed using an ideal HWP ($T=-c=1$, $\rho=s=0$ in Eq.~\eqref{eq:1lHWP}). 
With ideal and non-ideal maps in hand, we can calculate difference maps that quantify signal residuals due to HWP-related systematics. The resulting difference maps cover the entire sky, but we use a  \SI{40}{\percent} sky mask (\texttt{gal040}) \citep{Planck2015_diffuse_foreground} before calculating power spectra using {\tt PolSpice} \citep{Challinor2011}.

\begin{figure}
    \centering
    \includegraphics[width=0.48\textwidth]{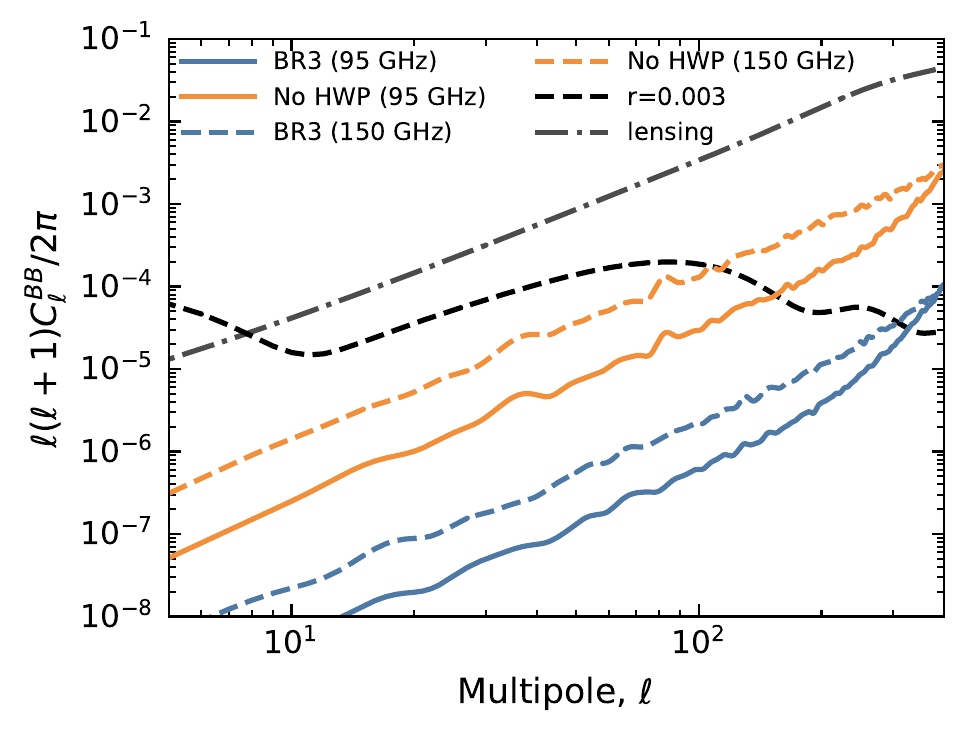}
    \caption{Residual \bmode power spectra  obtained by observing the CMB with the BR3 configurations presented in Table \ref{table:HWPconfig} (including the rotation angle offset optimized for the CMB). The beams are Gaussian. We omit the BR1 and BR5 HWP configurations since their residuals fall below the limits on the vertical axis. The no-HWP case is also shown (orange curves).}
    \label{fig:bmode_basic}
\end{figure}

% Calibration process
\subsection{Calibration}
\label{sec:results_cal}
To correct for the non-ideal polarization efficiency of each HWP model, we calibrate each map on a map obtained by scanning with an ideal HWP. This is performed using the $EE$ angular power spectrum at degree angular scales, $50\leq\ell\leq200$.  The choice of angular scales roughly coincides with the peak in the expected primordial gravitational wave power spectrum.  
Note that the calibration procedure could instead be performed using lab measurements or simulated HWP (and other optical component) material properties \citep{Pisano2006, Bryan2010a, Bryan2010b, Hill2016}. The $EE$ calibration approach uses the following factor: 
\begin{equation}
g = \frac{1}{151} \sum _{\ell = 50}^{200} \frac{C^{EE, \mathrm{ideal}}_{\ell}} {C_\ell^{EE}}\, ,
\end{equation}
where the denominator (numerator) is the $E$-mode power spectrum estimated from the output maps created with a non-ideal (ideal) HWP. The final difference maps are formed by subtracting the calibrated output of the non-ideal simulation from the ideal simulation's output:
\begin{equation}
\begin{pmatrix} Q \\ U \end{pmatrix}_\mrm{diff}= \begin{pmatrix} Q \\ U \end{pmatrix}_\mrm{ideal} -\sqrt{g} \begin{pmatrix} Q \\ U \end{pmatrix} \, .
\end{equation}
The residual \bmode power spectrum caused by the non-ideal HWP is then estimated from these calibrated difference maps. %Note that we have verified that results in the next section are unchanged if we base the above calibration method on the $B$-mode power spectrum instead. A $BB$ calibrator would be ideal, but would be unavailable in a realistic scenario. 

Finally, we divide out a beam window function to correct the power spectrum for the azimuthally symmetric part of the beam. This allows us to directly compare the residual to theory spectra. For each simulation we use a window function that corresponds to the averaged symmetric part of the input detector beams.

\subsection{Scanning with an ideal Gaussian beam}
\label{sec:results_basic}
We start by exploring effects that are purely caused by non-ideal HWPs. This is achieved by choosing a co-polar polarized and azimuthally symmetric Gaussian beam model, see e.g.\ \cite{Duivenvoorden2018}. Using this beam, we scan the CMB with the different HWP configurations; we summarise our results in Fig.~\ref{fig:bmode_basic}.  %shows the residual \bmode power spectra. 
We find that only the BR3 configuration shows an appreciable \bmode residual in this case. All three HWP configurations outperform the case without HWP modulation, which shows a relatively large white-noise spectrum caused by small conditioning problems in the map-making solution that are approximately uncorrelated between pixels.
It is instructive to determine which terms of the data model in Eqs.~\eqref{eq:blm_i}-\eqref{eq:blm_v} are causing the BR3 residual. It turns out that this spurious signal is due to $E\rightarrow B$ leakage from the $4\nu_{\alpha}$ terms, i.e.\ non-idealities in the inner two-by-two part of the HWP Mueller matrix. We have checked that the residual is not caused by $I\rightarrow (Q, U)$ leakage due to the $2 \nu_{\alpha}$ term in Eq.~\eqref{eq:blm_i} that couples the linearly polarized beam to the $I$ sky signal: we obtain virtually identical residuals when the input Stokes $I$ signal is artificially set to zero. The insignificance of the $2 \nu_{\alpha}$ term  can be attributed to the smallness of the $IQ$ and $IU$ elements in the HWP Mueller matrices (see Fig.~\ref{fig:MM_elms_alpha}), the lack of a strong atmospheric $I$ signal and, most importantly, the rather good conditioning of the map-making solution. Even without modification, the map-maker corresponding to Eq.~\eqref{eq:mapmaker} accurately distinguishes between time-ordered signal that is modulated at $2 \nu_{\alpha}$ and $4 \nu_{\alpha}$.
 
\begin{figure}
    \centering
    \includegraphics[width=0.48\textwidth]{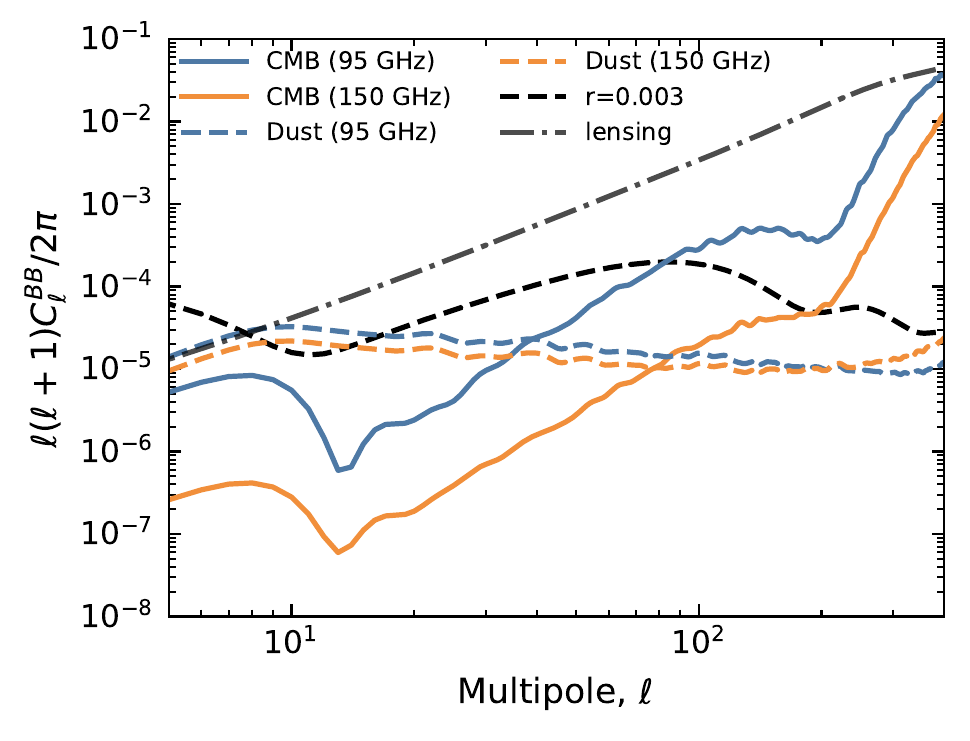}
    \caption{Residual \bmode power spectra generated when the CMB is observed using the BR3 HWP with a rotation angle offset optimized for the \texttt{PySM} Galactic dust model \texttt{d1} (solid curves) and vice versa (dashed curves).
    \label{fig:bmode_phase}}
\end{figure}

Using the same setup, we then explore the addition of a foreground component. Specifically, we simulate what happens when a map-maker that uses an HWP angle offset $\varphi$ (see Eq.~\eqref{eq:mapmaker}) that is optimized for the CMB encounters polarized signal from Galactic dust. Fig.~\ref{fig:bmode_phase} shows the \bmode residual for this hypothetical situation as well as for the opposite case in which the CMB is observed with $\varphi$ optimized for the SED of dust. %by observations of the CMB at 95 and \SI{150}{\giga\hertz} using the BR3 HWP that is optimized for the SED of Galactic dust. 
We again only show the BR3 HWP configuration.  The error in $\varphi$ causes $E \rightarrow B$ leakage: the residual clearly traces the shape of the input $E$-mode spectrum. The effect is identical to that of a systematic polarization angle calibration error. 
It can be seen that for both cases the residual is larger for \SI{95}{\giga\hertz} than for \SI{150}{\giga\hertz}. This is due to the fact that the optimal BR3 offset angle for dust in the \SI{95}{\giga\hertz} band differs from the optimal angle offset for the CMB by about \ang{0.4} while the difference at \SI{150}{\giga\hertz} is only half that.

From this section it becomes clear that in the presence of multiple sky components a single HWP offset angle $\varphi$ will not  effectively reduce $B$-mode residual caused by HWP non-idealities. The remaining spurious signal for the BR3 HWP configuration is at a level that would be unacceptable for upcoming $B$-mode experiments. A correction angle per sky component seems to be necessary. We further explore this point in the next section.

%\newpage
\begin{figure}
    \centering
    \includegraphics[width=0.48\textwidth]{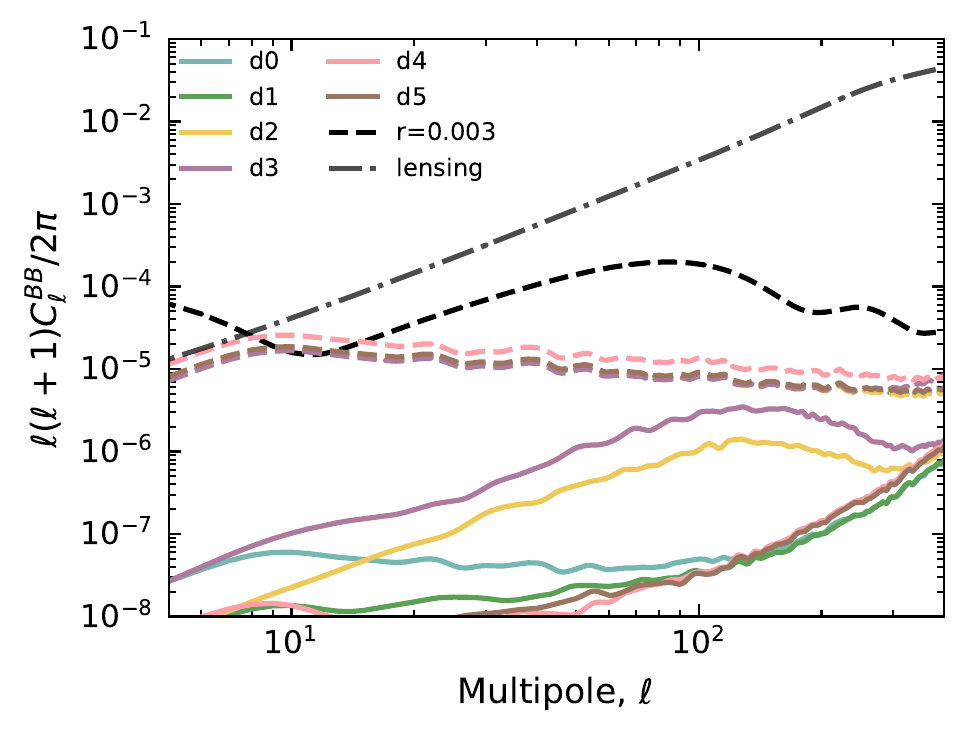}
    \caption{Residual \bmode power spectra for the different \texttt{PySM} Galactic dust models in the \SI{150}{\giga\hertz} frequency band scanned using the BR3 HWP configuration. The solid lines use a value of the HWP angle offset that is tailored to each dust model (the median of the distributions shown in Fig.~\ref{fig:dustanglehist}). The dashed colored lines use the median of the rotation angle offsets calculated for the case of an SED given by the combination of CMB and dust.  
    }   
    \label{fig:dspec_all}
    
\end{figure}

\begin{figure*}
    \centering
    \includegraphics[width=0.48\textwidth]{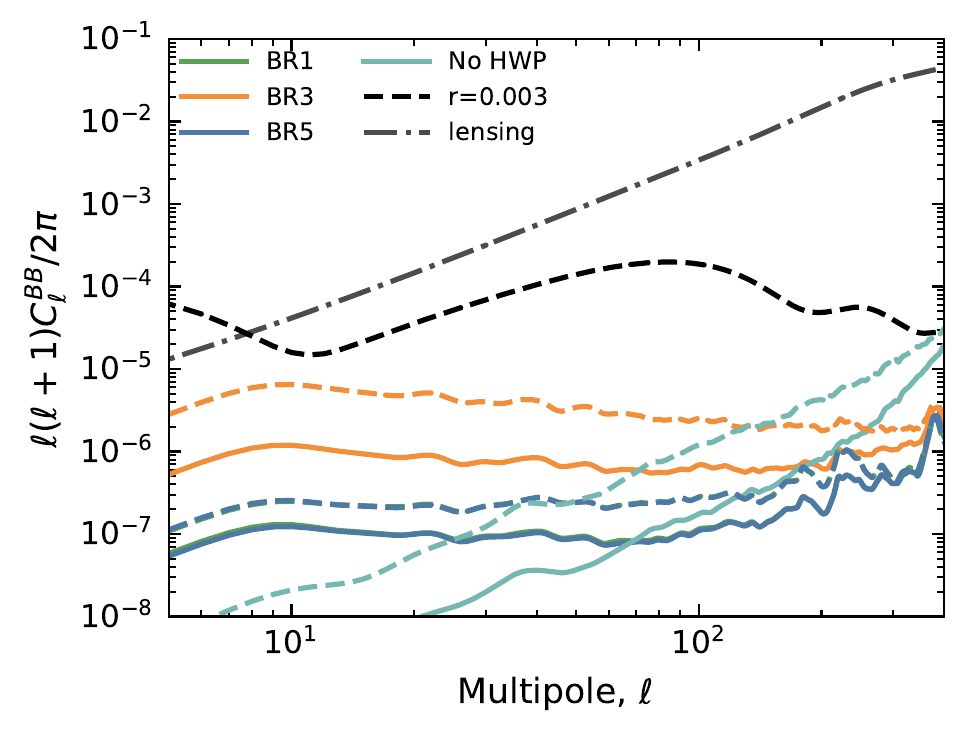}
     \includegraphics[width=0.48\textwidth]{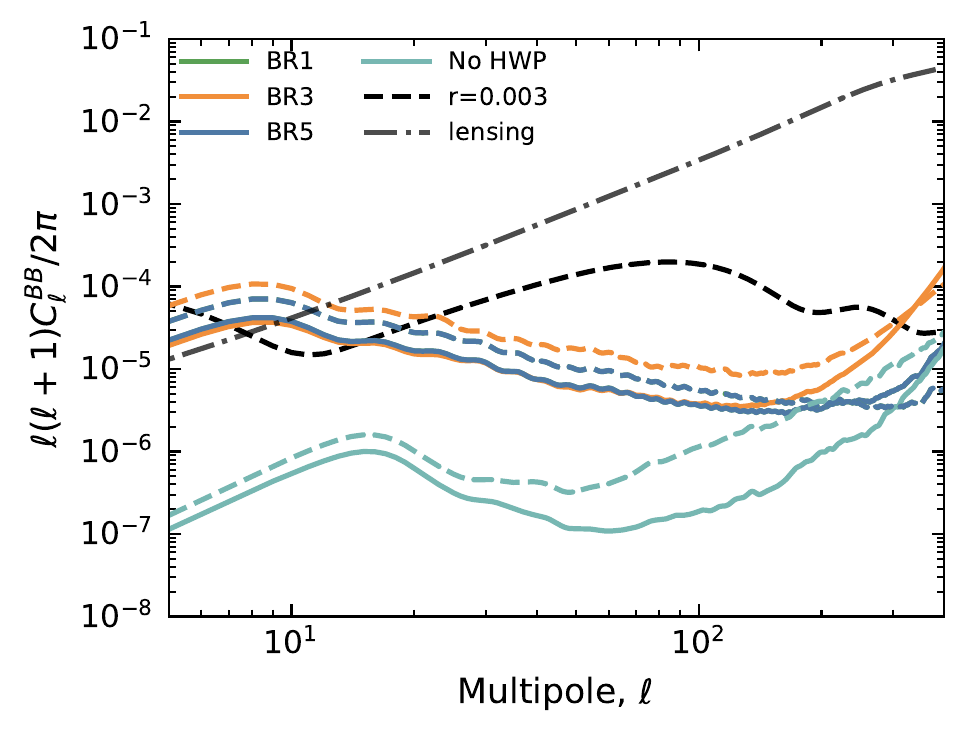}
    \caption{Left: Residual \bmode power spectra at \SI{95}{\giga\hertz} (solid lines) and \SI{150}{\giga\hertz} (dashed lines) derived from the band-averaged difference maps obtained by observing the \texttt{PySM} \texttt{d1} dust model using all of the HWP configurations presented in Table \ref{table:HWPconfig} for scans with a physical optics beam truncated at \ang{3}. Right: The same, but now observing the sky with a physical optics beam that extends to \ang{30} and therefore includes a higher contribution from sidelobes (see Fig.~\ref{fig:profile}). Note that the BR1 curves are almost completely hidden behind the BR5 curves. 
    } 
    \label{fig:bmode_residual_po}
\end{figure*}
\subsection{Foreground dependence}
\label{sec:fgdep}

To investigate how the HWP-induced systematics depend on foreground emission, we scan the different \texttt{PySM} Galactic dust models (\texttt{d0}-\texttt{d5}) with Gaussian beams (using the same setup as in the previous section). Data from the \textit{Planck} satellite have provided a wealth of information on  
Galactic dust emission, but there remains considerable uncertainty regarding both its frequency scaling and spatial variation \citep{Planck2018-11}. It is therefore natural to ask whether this uncertainty is large enough to impact the modelling of HWP systematics.  
We are particularly interested in seeing if spatial variation in the effective spectral index invalidates the use of a single HWP rotation angle offset. Recall that in  Fig.~\ref{fig:dustanglehist} the offset angles for the various {\tt PySM} dust models are compared to the offset angle determined for the simplest modified black-body model \texttt{d0}. The offset angle distributions of the more involved dust models are both biased from the \texttt{d0} value and show a dispersion. The model with the greatest dispersion (\texttt{d3}) predicts that a significant number of sky pixels will have an optimal offset angle that is more than \ang{0.1} away from the mean value for the BR3 HWP configuration.   

Fig.~\ref{fig:dspec_all} shows the effect of ignoring the spatial SED variations of the various \texttt{PySM} models. We scan the dust models using the BR3 HWP and correct for the HWP-induced rotation offset using an angle that corresponds to the mean of each distribution in Fig.~\ref{fig:dustanglehist}. As expected, we see that the \texttt{d2} and \texttt{d3} models, which both have a relatively large spread in spectral index over the sky, give the largest residuals. However, the amplitude of the spurious signal is still well below any detectable $B$-mode power spectrum amplitude. It thus seems that any realistic spatial variation in the dust SED can be safely ignored when determining the optimal HWP rotation angle correction for the dust component.

Similar to the previous section, we also explore the case in which a single angle calculated for the SED of the combination of CMB and dust is used to correct for the HWP-induced rotation angle. These residuals are given by the dashed lines in Fig.~\ref{fig:dspec_all}. We again see that this choice of correction angle would produce significant residual and we see that this results is insensitive to the choice of dust model.

\subsection{Scanning with a non-ideal beam}

The simulation framework presented in this paper enables studies of the complicated interplay between non-ideal HWPs and non-ideal beams. For this purpose, we can use physical optics (PO) simulations that include extended beam sidelobes with non-negligible cross-polar response; features that could be present in an optical configuration shown in Fig.~\ref{fig:tel_des}. The azimuthally averaged beam profiles for the Stokes $Q$ and $U$ beams of a representative beam used in this analysis are shown in Fig.\ \ref{fig:profile}. We study two cases, one where we apodize the beam maps at \ang{3} away from the beam center (no far-sidelobes) and one where we extend our beam maps out to \ang{30} (with far-sidelobes). In order to focus on effects from the interplay between the beam and the HWP, we calculate difference maps by subtracting a map generated using the same beam model but with an ideal HWP.

Fig.~\ref{fig:bmode_residual_po} shows the resultant \bmode residuals; the input sky is the \texttt{d1} dust model, the amplitude of the curves should be compared to the solid \texttt{d1} curve in Fig.~\ref{fig:dspec_all}.  
The effect of the more complex beam model is twofold. The increased solid angle of the beam, i.e.\ the sidelobe, brings in $E$-mode dust signal from behind the Galactic mask. Given that we use a  correction for the HWP rotation angle offset $\varphi$ that has been calculated for unmasked pixels, the correction that we apply is not quite appropriate for this extra signal. The result is $E \rightarrow B$ leakage close to the edges of the mask. The second, more significant, effect is due to the cross-polar beam. This is especially obvious in the right panel of Fig.~\ref{fig:bmode_residual_po} that was made  with the beam model that extends out to \ang{30} and includes a relatively large cross-polar component. The impact of the cross-polar beam can be understood as an $\ell$-dependent polarization rotation that, given the shape of the cross-polar component in Fig.~\ref{fig:profile}, is larger at lower $\ell$. One might wonder why the resulting $E \rightarrow B$ leakage is not canceled in our setup when we subtract the ideal-HWP maps that were created using the same cross-polar beam. The reason is that the dominant HWP non-ideality couples directly to the cross-polar beam component: the two effects are not additive but multiplicative. This can be seen in the third line of Eq.~\eqref{eq:blm_p}: the dominant $4 \nu_{\alpha}$ term of the data model contains a term proportional to $\smash{\widetilde{U}^{(0)}_{\mathrm{b}}C_{P^* P}}$, i.e.\ the product of the cross-polar beam and the $P^* P$ component of the HWP Mueller matrix in  Eq.~\eqref{eq:complex_elements}. Roughly speaking, the difference maps used to create the spectra in Fig.~\eqref{fig:bmode_residual_po} are thus proportional to the cross-polar beam times $(1 - C_{P^* P})$, the deviation from the ideal HWP Mueller element. The outcome is $E \rightarrow B$ leakage from the HWP non-ideality that is modulated by the cross-polar beam, resulting in the leaking of a redder version of the original $E$-mode dust spectrum to the $B$-mode spectrum, as can be observed in the right panel of Fig.~\eqref{fig:bmode_residual_po}.

\subsection{Polarization sensitivity}
\label{sec:results_sensitivity}
Given the results that we have discussed so far, there does not seem to be much difference between the BR1 and BR5 performance. Both outperform the BR3 HWP configuration in all the tests we presented and in Fig.~\ref{fig:bmode_residual_po} the BR1 and BR5 curves overlap almost perfectly. However, the calibration process that we described in Sec.~\ref{sec:results_cal} masks the fact that the BR5 configuration has much greater polarization modulation efficiency than the BR1 configuration. For example, in the case when we scan the CMB with a Gaussian beam (see Sec.~\ref{sec:results_basic}, Fig.~\ref{fig:bmode_basic}), we find that the calibration coefficients based on the $E$-mode power spectrum are 1.44, 1.10, 1.09, and 1.00 for the BR1, BR3, BR5, and no HWP configurations, respectively. In comparison, the calibration procedure that uses the temperature power spectrum gives 1.04, 1.05, 1.08, and 1.00, for the BR1, BR3, BR5, and no-HWP configurations, respectively. This shows that even though the BR5 configuration has lower optical efficiency because of the larger number of optical elements, and therefore a greater number of both loss and reflection mechanisms, its polarization modulation efficiency, and therefore sensitivity, is approximately \SI{15}{\percent} higher than that of the BR1 configuration when integrated over the \SI[number-unit-product=\text{-}]{95}{\giga\hertz} band.
\section{Conclusions}

We formulated an extension of the harmonic beam convolution algorithm originally described by \cite{wandelt_2001} that adds the capability of simulating systematics due to non-ideal half-wave plates (HWPs). The generalized algorithm allows for numerically efficient generation of simulated time-domain data that include spurious signal from non-ideal HWPs  and asymmetric and/or non-trivially polarized beams.  
Such time-domain simulations are a crucial part of  ``end-to-end'' analysis pipeline validation efforts for CMB experiments. As multiple current and upcoming CMB instruments employ HWPs, it is timely to include the associated non-idealities in our simulations. The new simulator also allows us to investigate the importance of HWP-related systematics, some of which we have investigated in this paper. The extended algorithm is implemented as part of the publicly available \texttt{beamconv} code, which has also been used to derive the results in this paper.

For our investigation into HWP systematics, we included three different HWP configurations: a 1-, 3-, and 5-layer model. With this selection, we simulated data for a representative CMB $B$-mode satellite experiment that employs a spinning HWP as polarization modulator. Particular attention was paid to the frequency dependence of the system. Our simulated experiment employs dichroic detectors and is thus especially sensitive to frequency dependent  HWP systematics given the wide frequency band of the detectors. 

We find that the choice of HWP configuration significantly impacts the $B$-mode reconstruction fidelity. In particular, the 3-layer HWP that we study comes with a significant frequency dependent rotation angle offset, which, if not corrected for, acts as a polarization angle offset that leaks $E$-mode to $B$-mode polarization by an amount that would be problematic for an experiment aiming to constrain the tensor-to-scalar ratio $r$ to a level of $r<0.003$. Correcting for the rotation offset requires a correcting HWP angle offset $\varphi$ that is dependent on the SED of the observed signal; we demonstrate that $\varphi$ varies significantly between the CMB signal and the Galactic dust signal. This introduces a challenge for the standard CMB data analysis paradigm, which aims to compress an experiment's time-ordered data into unbiased sky maps before component separation and cosmological analysis is performed. During this
map-making procedure one typically has no knowledge of the relative contribution of each sky component to the time-ordered data. As a result, the map-making procedure can only be given a single $\varphi$ angle, based on some combination of the optimal $\varphi$ of each of the sky components, which will necessary lead to biased maps. 
Parametric algorithms for component separation, starting from a prior on the SEDs of the various sky components, could use $\varphi$ as a parameter per sky component and forward propagate the polarization rotation. Such algorithms might attempt to divine the $\varphi$ angles from the observed amount of $EB$ signal in the non-component separated maps, as no significant $EB$ power has until now been observed for either dust or the CMB \citep{Planck2018-11}.

In light of HWP rotation angle offsets that vary between sky components, we investigate how well one would need to know the SED of polarized Galactic dust when modelling the angle offset of this component. We find that the current understanding of the dust SED will likely suffice for this procedure. We determine offset angles for a range of different dust models and find that the resulting angles vary by an insignificant amount. Spatial variations in the dust SED also seem to be of relatively minor importance.

Finally, we leverage the potential of the new code by simulating data using non-ideal HWPs and non-ideal instrumental beams. We point out that there exist an interplay between the cross-polar component of the beam and certain HWP non-idealities. We find significant $B$-mode residual for all three HWP configurations when this interplay is not modelled correctly. We can conclude that a thorough understanding of the instrumental beam will be necessary for future  experiments attempting to model or correct for HWP non-idealities.

\label{sec:Conclusion}

\section*{Acknowledgements}
We are grateful to 
Aurelien Fraisse, Brandon Hensley, Jo Dunkley, Tomotake Matsumura, and Hans Kristian Eriksen
for helpful comments. Computations have been performed at the Owl Cluster
funded by the University of Oslo and the Research Council of Norway through grant 250672. JEG acknowledges support from the Swedish National Space Agency (SNSA/Rymdstyrelsen) and the Swedish Research Council (Reg. no. 2019-03959). Some of the results in
this paper have been derived using the HEALPix \citep{Healpix2005} package.
%The Acknowledgements section is not numbered. Here you can thank helpful
%colleagues, acknowledge funding agencies, telescopes and facilities used etc.

%%%%%%%%%%%%%%%%%%%%%%%%%%%%%%%%%%%%%%%%%%%%%%%%%%


\bibliographystyle{mnras}
\bibliography{hwp}

%%%%%%%%%%%%%%%%%%%%%%%%%%%%%%%%%%%%%%%%%%%%%%%%%%

%%%%%%%%%%%%%%%%% APPENDICES %%%%%%%%%%%%%%%%%%%%%

\appendix

\section{Expanding on the Mathematical Framework}
\label{sec:appendixmath}
The aim of this appendix is to give a more exhaustive explanation of the mathematical framework used in  Sec.~\ref{sec:MathFramework}. In particular, we will derive the harmonic-domain version of the data model of Eq.~\eqref{eq:new_data_model} and derive the harmonic coefficients in Eqs.~\eqref{eq:blm_i}-\eqref{eq:blm_v}.

We express the data model in terms of the Stokes parameters of the instrument and the sky by inserting Eq.~\eqref{eq:I_det} in Eq.~\eqref{eq:data_model_gen}:
\begin{align} \label{eq:data_model_stokes}
\begin{split}
d_t = \int \mathrm{d}\nu \, F(\nu) \int \mathrm{d}\Omega (\hat{\bm{n}}) \Big( &I \widetilde{I}^{\,(t)}_{\mathrm{i}} + Q \widetilde{Q}^{\,(t)}_{\mathrm{i}} \\&+ U \widetilde{U}^{\,(t)}_{\mathrm{i}} 
 + V\widetilde{V}^{\,(t)}_{\mathrm{i}} \Big) (\hat{\mathbfit{n}}, \nu) \, .
\end{split}
\end{align}
Note that we omit the noise term for brevity. The instrumental Stokes parameters in the above equation are defined in a basis fixed to the sky and thus change continuously as the telescope scans over the sky. The transformation between sky and instrument coordinate frame is given by Eq.~\eqref{eq:instr2sky}. In this derivation we will however first express the data model in the harmonic domain before performing the transformation. 

By working in the harmonic domain we can make use of the fact that a generic set of spin-weighted spherical harmonic coefficients $\smash{f^{(0)}_{\ell m}}$ defined with respect to the coordinate basis fixed to the instrument transform as follows:
\begin{align}\label{eq:alm_trans}
f^{(0)}_{\ell m} \mapsto f^{(t)}_{\ell m} = \sqrt{\frac{4 \pi}{2 \ell + 1}} \sum_{s=-\ell}^{\ell} f^{(0)}_{\ell s}  \, {}_{s} Y_{\ell -m}(\theta_t, \phi_t) e^{-\mathrm{i} s \psi_t} \, ,
\end{align}
when we instead define the coefficients with respect to the coordinate frame fixed relative to the sky. 
Here, $\psi_i$, $\theta_i$, and $\phi_i$ are the 3 Euler angles that describe~$\mathbfss{R}_t$, the rotation between the two frames, and ${}_{s}Y_{\ell m}$ is a spin-$s$ spherical harmonic \citep{goldberg_1967, newman_1966}. 

To make use of Eq.~\eqref{eq:alm_trans} it is necessary to know the spin-weighted spherical harmonic coefficients for each of the different Stokes parameters in Eq.~\eqref{eq:data_model_stokes}. Using the transformation rule for the density matrix in Eq.~\eqref{eq:instr2sky}, we may illustrate why $\smash{\widetilde{I}^{\,(t)}_{\mathrm{i}}}$, and $\smash{\widetilde{V}^{\,(t)}_{\mathrm{i}}}$ should be expanded into regular (spin-0) spherical harmonics and why
\begin{align}
\widetilde{P}^{\,(t)}_{\mathrm{i}} = \widetilde{Q}^{\,(t)}_{\mathrm{i}} + \mathrm{i} \widetilde{U}^{\,(t)}_{\mathrm{i}} \, ,
\end{align}
ought to be expanded in spin-$2$ spherical harmonics. We note that the $\bm{\Lambda}$ matrices in Eq.~\eqref{eq:instr2sky} generally depend on the $\psi_t$, $\theta_t$, and $\phi_t$ angles that describe~$\mathbfss{R}_t$ but that in the case where $\mathbfss{R}_t$ describes a right-handed rotation around $\hat{\mathbfit{n}}$ by an angle $\psi_t$ the matrices are simply given by
\begin{align}
\Lambda_{i}^{\phantom{a}j} \left(\mathbfss{R}_{\hat{\mathbfit{n}}}(\psi_t)\right) = \begin{pmatrix} \cos \psi_t & \sin \psi_t \\ -\sin \psi_t & \cos \psi_t \end{pmatrix} \, .
\end{align}
It is straightforward to check that when this specific rotation is applied to $\smash{\mathbfss{W}_{\mathrm{instr}}^{(t)} 
}$, % the elements of the anti-symmetric part of $\smash{\mathbfss{W}_{\mathrm{instr}}^{(t)}}$, 
the $\smash{\widetilde{I}^{\,(t)}_{\mathrm{i}}}$ and $\smash{\widetilde{V}^{\,(t)}_{\mathrm{i}}}$ elements remain invariant, while the elements of the symmetric trace-free part, $\smash{\widetilde{Q}^{\,(t)}_{\mathrm{i}}}$ and $\smash{\widetilde{U}^{\,(t)}_{\mathrm{i}}}$, transform as a spin-$2$ field:
\begin{align}\label{eq:spin2}
\big(\widetilde{Q}^{\,(t)}_{\mathrm{i}} \pm \mathrm{i} \widetilde{U}^{\,(t)}_{\mathrm{i}}\big) (\hat{\mathbfit{n}}) \mapsto \mathrm{e}^{\mp 2 \mathrm{i} \psi_t}  \big(\widetilde{Q}^{\,(t)}_{\mathrm{i}} \pm \mathrm{i} \widetilde{U}^{\,(t)}_{\mathrm{i}}\big)(\hat{\mathbfit{n}}) \, .
\end{align}

We now expand the instrumental Stokes parameters into the appropriate spin-weighted spherical harmonics:
\begingroup
\allowdisplaybreaks
\begin{align}
\widetilde{I}^{\,(t)}_{\mathrm{i}}\!(\hat{\mathbfit{n}}, \nu,  \alpha_t) &= \sum_{\ell=0}^{\ell_{\mathrm{max}}} \sum_{m=-\ell}^{\ell} b^{\widetilde{I}^{\,(t)}_{\mathrm{i}}}_{\ell m}(\nu, \alpha_t) Y_{\ell m}(\hat{\mathbfit{n}}) \, ,\label{eq:I_beam_decomp} \\
\widetilde{P}^{\,(t)}_{\mathrm{i}}\,(\hat{\mathbfit{n}}, \nu,  \alpha_t)  &= \sum_{\ell=2}^{\ell_{\mathrm{max}}} \sum_{m=-\ell}^{\ell} {}_{2}b^{\widetilde{P}^{\,(t)}_{\mathrm{i}}}_{\ell m}(\nu, \alpha_t) {}_{2}Y_{\ell m}(\hat{\mathbfit{n}}) \, , \\
\widetilde{V}^{\,(t)}_{\mathrm{i}}\,(\hat{\mathbfit{n}}, \nu,  \alpha_t)  &= \sum_{\ell=0}^{\ell_{\mathrm{max}}} \sum_{m=-\ell}^{\ell} b^{\widetilde{V}^{\,(t)}_{\mathrm{i}}}_{\ell m}(\nu, \alpha_t) Y_{\ell m}(\hat{\mathbfit{n}}) \, .
\begin{split}
\end{split}
\end{align}
\endgroup
The Stokes parameters of the sky are expanded in a similar manner: 
\begingroup
\allowdisplaybreaks
\begin{align}
I(\hat{\mathbfit{n}}, \nu) &= \sum_{\ell=0}^{\ell_{\mathrm{max}}} \sum_{m=-\ell}^{\ell} a^{I}_{\ell m}(\nu) Y_{\ell m}(\hat{\mathbfit{n}}) \, , \\
P(\hat{\mathbfit{n}}, \nu)  &= \sum_{\ell=2}^{\ell_{\mathrm{max}}} \sum_{m=-\ell}^{\ell} {}_{2}a^{P}_{\ell m}(\nu) {}_{2}Y_{\ell m}(\hat{\mathbfit{n}}) \, , \\
V(\hat{\mathbfit{n}}, \nu)  &= \sum_{\ell=0}^{\ell_{\mathrm{max}}} \sum_{m=-\ell}^{\ell} a^{V}_{\ell m}(\nu) Y_{\ell m}(\hat{\mathbfit{n}}) \, , \label{eq:v_sky_decomp}
\begin{split}
\end{split}
\end{align}
\endgroup
where we have used the following definition:
\begin{align}
P = Q + \mathrm{i} U \, .
\end{align}

We insert Eqs.~\eqref{eq:I_beam_decomp}-\eqref{eq:v_sky_decomp} into Eq.~\eqref{eq:data_model_stokes} to produce the following version of the data model:
\begingroup
\allowdisplaybreaks
\begin{align}\label{eq:data_model_harm}
\begin{split}
&d_t = \int \mathrm{d}\nu F(\nu) \sum_{\ell=0}^{\ell_{\mathrm{max}}} \sum_{m=-\ell}^{\ell} \Big\{ \big[b^{\widetilde{I}^{\,(t)}_{\mathrm{i}}}_{\ell m}(\nu, \alpha_t)\big]^* \, a^{I}_{\ell m}(\nu) \\
&\, \, \, \, +  \mathrm{Re}\Big( \! \big[{}_{2}b^{\widetilde{P}^{\,(t)}_{\mathrm{i}}}_{\ell s}\!\!(\nu, \alpha_t)\big]^* {}_{2}a^{P}_{\ell m} (\nu) \Big) + \big[b^{\widetilde{V}^{\,(t)}_{\mathrm{i}}}_{\ell s}\!\!(\nu, \alpha_t)\big]^* a^{V}_{\ell m}(\nu)\Big\} \, .
\end{split}
\end{align}
\endgroup
To obtain this expression, we have made use of the orthogonality of the spin-weighted spherical harmonics:
\begin{align}
\int_{S^2} \mathrm{d}\Omega(\hat{\mathbfit{n}}) {}_sY_{\ell m}(\hat{\mathbfit{n}}) {}_sY^*_{\ell' m'}(\hat{\mathbfit{n}}) = \delta_{\ell, \ell'} \delta_{m, m'} \, .
\end{align}
Note that the $\smash{b^{\widetilde{I}_t}_{\ell m}}$, $\smash{{}_{2}b^{\widetilde{P}_t}_{\ell m}}$, and $\smash{b^{\widetilde{V}_t}_{\ell m}}$ coefficients in Eq.~\eqref{eq:data_model_harm} are still defined on the basis fixed to the sky, so they are  time dependent (they change as the telescope scans over the sky). We may now use Eq.~\eqref{eq:alm_trans} to relate these time-varying  coefficients to the $\smash{b^{\widetilde{I}}_{\ell s}}$, $\smash{{}_{\pm2}b^{\widetilde{P}}_{\ell s}}$, and $\smash{b^{\widetilde{V}}_{\ell s}}$ coefficients in Eq.~\eqref{eq:new_data_model} that are defined with respect to the coordinate frame fixed to the instrument. %This transforming from the coordinate basis fixed to the sky to system fixed to the instrument. 
Under the rotation $\mathbfss{R}_t$ the following relationships hold:
\begingroup
\allowdisplaybreaks
\begin{align}
\begin{split}
b^{\widetilde{I}^{\,(0)}_{\mathrm{i}}}_{\ell m}(\alpha_t) \mapsto& \, b^{\widetilde{I}^{\,(t)}_{\mathrm{i}}}_{\ell m}(\alpha_t) \\
&= q_{\ell} \sum_{s=-\ell}^{\ell} b^{\widetilde{I}^{\,(0)}_{\mathrm{i}}}_{\ell s}(\alpha_t)  \, {}_{s} Y_{\ell -m}(\theta_t, \phi_t) e^{-\mathrm{i} s \psi_t} \, , 
\end{split} \\
\begin{split}
{}_{2}b^{\widetilde{P}^{\,(0)}_{\mathrm{i}}}_{\ell m}(\alpha_t) \mapsto& \, {}_{2}b^{\widetilde{P}^{\,(t)}_{\mathrm{i}}}_{\ell m}(\alpha_t) \\
&= q_{\ell} \sum_{s=-\ell}^{\ell} {}_{2}b^{\widetilde{P}^{\,(0)}_{\mathrm{i}}}_{\ell s}(\alpha_t)  \, {}_{s} Y_{\ell -m}(\theta_t, \phi_t) e^{-\mathrm{i} s \psi_t} \, , 
\end{split} \\
\begin{split}
b^{\widetilde{V}^{\,(0)}_{\mathrm{i}}}_{\ell m}(\alpha_t) \mapsto& \,  b^{\widetilde{V}^{\,(t)}_{\mathrm{i}}}_{\ell m}(\alpha_t) \\
&= q_{\ell} \sum_{s=-\ell}^{\ell} b^{\widetilde{V}^{\,(0)}_{\mathrm{i}}}_{\ell s}(\alpha_t)  \, {}_{s} Y_{\ell -m}(\theta_t, \phi_t) e^{-\mathrm{i} s \psi_t} \, ,
\end{split}
\end{align}
\endgroup
where we have defined the shorthand:
\begin{align}
q_{\ell} = \sqrt{\frac{4 \pi}{2 \ell + 1}} \, .
\end{align}
Inserting the above into Eq.~\eqref{eq:data_model_harm} yields the final expression for the data model in  Eq.~\eqref{eq:new_data_model}. 
%As a recap .. Recall that we define the Euler angles such that the harmonic coefficients on the right hand side correspond to the beam centred on the north pole of the $(\theta, \phi)$ coordinate system. 

To derive the harmonic coefficients in Eqs.~\eqref{eq:blm_i}-\eqref{eq:blm_v} we need to compute the instrumental Stokes parameters in the coordinate frame fixed to the instrument. We make use of Eq.~\eqref{eq:fact_beam_hwp} that expresses these parameters in terms of a Stokes vector representing the beam and the HWP Mueller matrix, rotated by an angle $\alpha_t$:
\begin{align}\label{eq:stokes_instrument_app}
\mathbfit{S}^{(0)\mathsf{T}}_{\mrm{instr}}(\hat{\mathbfit{n}}, \nu, \alpha_t) = \mathbfit{S}^{(0) \mathsf{T}}_{\mrm{beam}} (\hat{\mathbfit{n}}, \nu) \, \mathbfss{M}_{\mrm{HWP}} (\nu, \alpha_t) \, .
\end{align}
The instrumental Stokes vector contains the same information as the instrumental density matrix $\mathbfss{W}^{(0)}_{\mathrm{instr}}$ in Eq.~\eqref{eq:instr2sky}. We may use Eq.~\eqref{eq:stokes2rho} to transform the between density matrix and Stokes vector using the following Pauli matrices:
\begin{align}
    (\sigma_0)_{ij} &= \begin{pmatrix} 1 & 0 \\ 0 & \sin^2\theta \end{pmatrix} \, , \label{eq:pauli_0}\\
    (\sigma_3)_{ij} &= \begin{pmatrix} 1 & 0 \\ 0 & -\sin^2\theta \end{pmatrix} \, , \\
    (\sigma_1)_{ij} &= \begin{pmatrix} 0 & \sin\theta \\ \sin\theta & 0 \end{pmatrix} \, , \\
    (\sigma_2)_{ij} &= \begin{pmatrix} 0 & -\mathrm{i}\sin\theta \\ \mathrm{i}\sin\theta & 0 \end{pmatrix} \, .\label{eq:pauli_2}
\end{align}
The additional factors of $\sin \theta$ compared to the standard Pauli matrices are a consequence of the metric of the assumed spherical coordinates: $g_{ij} = \mathrm{diag}(1, \sin^2 \theta )$.

We start by rewriting Eq.~\eqref{eq:stokes_instrument_app} as follows:
\begin{align}
\mathbfit{S}^{(0) \mathsf{T}}_{\mrm{instr}}(\hat{\mathbfit{n}}, \nu, \alpha_t) \mathbfss{T}^{\dagger} = \mathbfit{S}^{(0)\mathsf{T}}_{\mrm{beam}} (\hat{\mathbfit{n}}, \nu) \mathbfss{T}^{\dagger} \mathbfss{T} \mathbfss{M}_{\mrm{HWP}} (\nu, \alpha_t)\mathbfss{T}^{\dagger} \, ,
\end{align}
where we have introduced the following complex transformation matrix:
\begin{align}
\mathbfss{T} = \begin{pmatrix}
1 & 0 & 0 & 0 \\ 
0 & \frac{1}{\sqrt{2}} & \frac{\mathrm{i}}{\sqrt{2}} & 0 \\
0 & \frac{1}{\sqrt{2}} & \frac{-\mathrm{i}}{\sqrt{2}} & 0 \\
0 & 0 & 0 & 1
\end{pmatrix} \, ,
\end{align}
that should be understood as transforming the real Stokes parameter basis to a complex basis spanned by $I$, $(Q + \mathrm{i}U)/\sqrt{2}$, $(Q - \mathrm{i}U)/\sqrt{2}$ and $V$. Note that $\mathbfss{T}$ is unitary:
\begin{align}
\mathbfss{T}^{\dagger} \mathbfss{T} = \mathbfss{T} \mathbfss{T}^{\dagger} = \mathbfss{1} \, .
\end{align}
Next, we factor the rotated HWP Mueller matrix into the unrotated matrix and two Mueller rotation matrices:
\begin{align}
\mathbfss{M}_{\mathrm{HWP}}(\alpha) = \mathbfss{M}_{\alpha}^{\mathsf{T}} \mathbfss{M}_{\mathrm{HWP}} \mathbfss{M}_{\alpha} \, ,
\end{align}
with:
\begin{align}
\mathbfss{M}_{\alpha} = \begin{pmatrix}
1 & 0 & 0 & 0 \\ 
0 & \cos 2\alpha & \sin 2 \alpha & 0 \\
0 & -\sin 2 \alpha & \cos 2 \alpha & 0 \\
0 & 0 & 0 & 1
\end{pmatrix} \, .
\end{align}
Note that the $\mathbfss{T}$ matrix diagonalizes the rotation matrix:
\begin{align}
\mathbfss{T} \mathbfss{M}_{\alpha} \mathbfss{T}^{\dagger} = \begin{pmatrix}
1 & 0 & 0 & 0 \\ 
0 & \mathrm{e}^{-2\mathrm{i}\alpha} & 0 & 0 \\
0 & 0 & \mathrm{e}^{2\mathrm{i}\alpha} & 0 \\
0 & 0 & 0 & 1
\end{pmatrix} \, .
\end{align}
Putting everything together yields:
\begin{align}
\begin{split}
&\mathbfit{S}^{(0) \mathsf{T}}_{\mrm{instr}}(\hat{\mathbfit{n}}, \nu, \alpha_t) \mathbfss{T}^{\dagger} = \\ & \quad \quad \mathbfit{S}^{(0) \mathsf{T}}_{\mrm{beam}}(\hat{\mathbfit{n}}, \nu) \mathbfss{T}^{\dagger} \mathbfss{T} \mathbfss{M}_{\alpha}^{\mathsf{T}} \mathbfss{T}^{\dagger} \mathbfss{T} \mathbfss{M}_{\mathrm{HWP}}(\nu)  \mathbfss{T}^{\dagger} \mathbfss{T} \mathbfss{M}_{\alpha} \mathbfss{T}^{\dagger} \, .
\end{split}
\end{align}
Evaluating this expression provides us with the instrumental Stokes parameters in terms of the beam Stokes parameters and the HWP:
\begingroup
\begin{align}
\begin{split}
\widetilde{I}^{\,(0)}_{\mathrm{i}}(\hat{\mathbfit{n}}, \alpha_t, \nu) &= \widetilde{I}^{\,(0)}_{\mathrm{b}} (\hat{\mathbfit{n}}, \nu)  C_{IV}(\nu)  + \widetilde{V}^{(0)}_{\mathrm{b}} \! (\hat{\mathbfit{n}}, \nu) C_{VV}(\nu) \\
&\quad +  \sqrt{2} \mathrm{Re} \Big( \widetilde{P}^{\,(0)}_{\mathrm{b}} \! (\hat{\mathbfit{n}}, \nu)  C_{P^* V}(\nu)\mathrm{e}^{-2\mathrm{i}\alpha} \Big) \label{eq:stokes_instr_i}\, ,
\end{split}  \\
\begin{split}
\widetilde{P}^{(0)}_{\mathrm{i}} (\hat{\mathbfit{n}}, \alpha, \nu) &= \widetilde{I}^{\,(0)}_{\mathrm{b}} (\hat{\mathbfit{n}}, \nu)  C_{IP}(\nu) \sqrt{2} \, \mathrm{e}^{-2\mathrm{i}\alpha} \\
&\quad +  \widetilde{V}^{(0)}_{\mathrm{b}} \!(\hat{\mathbfit{n}}, \nu)  C_{VP}(\nu) \sqrt{2}  \mathrm{e}^{-2\mathrm{i}\alpha}  \\
&\quad   +  \widetilde{P}^{(0)}_{\mathrm{b}} \!(\hat{\mathbfit{n}}, \nu)  C_{P^* P} (\nu) \mathrm{e}^{-4\mathrm{i}\alpha} \\
&\quad+ \widetilde{P}^{(0)*}_{\mathrm{b}} (\hat{\mathbfit{n}}, \nu)  C_{P P}(\nu) \, ,
\end{split}   \\
\begin{split}
\widetilde{V}^{(0)}_{\mathrm{i}} (\hat{\mathbfit{n}}, \alpha, \nu) &= \widetilde{I}^{\,(0)}_{\mathrm{b}} \!(\hat{\mathbfit{n}}, \nu)  C_{IV}(\nu)  + \widetilde{V}^{(0)}_{\mathrm{b}} \! (\hat{\mathbfit{n}}, \nu) C_{VV}(\nu) \\
&\quad+  \sqrt{2} \mathrm{Re} \Big( \widetilde{P}^{\,(0)}_{\mathrm{b}} \! (\hat{\mathbfit{n}}, \nu)  C_{P^* V}(\nu)\mathrm{e}^{-2\mathrm{i}\alpha} \Big)\label{eq:stokes_instr_v} \, , 
\end{split}
\end{align}
\endgroup
where:
\begin{align}
\widetilde{P}^{\,(0)}_{\mathrm{i}} = \widetilde{Q}^{\,(0)}_{\mathrm{i}} + \mathrm{i} \widetilde{U}^{\,(0)}_{\mathrm{i}} \, ,
\end{align}
and where we have used the following shorthand for the unrotated HWP Mueller matrix expressed in the complex basis:
\begin{align}
\mathbfss{C} = \mathbfss{T} \mathbfss{M}_{\mathrm{HWP}}  \mathbfss{T}^{\dagger} \, ,
\end{align}
that, in terms of the original HWP Mueller matrix elements, is given by:
\begin{align}\begin{split}     
\mathbfss{C} = &\left(
\begin{matrix}
M_{II} & \frac{M_{IQ}- \mathrm{i} M_{IU}}{\sqrt{2}} \\
\frac{M_{QI}+\mathrm{i}M_{UI}}{\sqrt{2}} & \frac{M_{QQ}  + M_{UU} - \mathrm{i} (M_{QU} - M_{UQ})}{2} \\
\frac{M_{QI}-\mathrm{i}M_{UI}}{\sqrt{2}} & \frac{M_{QQ}  - M_{UU} - \mathrm{i} (M_{QU} + M_{UQ})}{2} \\
M_{VI} & \frac{M_{VQ}-\mathrm{i}M_{VU}}{\sqrt{2}} 
\end{matrix}
\right.\\
&\quad \quad \left.
\begin{matrix}
\frac{M_{IQ}+\mathrm{i}M_{IU}}{\sqrt{2}} & M_{IV}\\
\frac{M_{QQ}  - M_{UU} + \mathrm{i} (M_{QU} + M_{UQ})}{2} & \frac{M_{QV}+\mathrm{i}M_{UV}}{\sqrt{2}}\\ 
\frac{M_{QQ}  + M_{UU} + \mathrm{i} (M_{QU} - M_{UQ})}{2} & \frac{M_{QV}-\mathrm{i}M_{UV}}{\sqrt{2}}\\
\frac{(M_{VQ}+\mathrm{i}M_{VU})}{\sqrt{2}} & M_{VV} 
\end{matrix}
\right) \, .
\end{split}
\label{eq:complex_elements}
\end{align}
Finally, we plug the instrumental Stokes parameters in Eqs.~\eqref{eq:stokes_instr_i}-\eqref{eq:stokes_instr_v} into the transformations below:
\begingroup
\allowdisplaybreaks
\begin{align}
\begin{split}
b^{\widetilde{I}^{\,(0)}_{\mathrm{i}}}_{\ell m}(\nu) &= \int_{S^2} \mathrm{d}\Omega(\hat{\mathbfit{n}}) \widetilde{I}^{\,(0)}_{\mathrm{i}}(\hat{\mathbfit{n}}, \alpha_t, \nu) Y_{\ell m}^* (\hat{\mathbfit{n}})  \, ,
\end{split} \\
\begin{split}
{}_{2}b^{\widetilde{P}^{\,(0)}_{\mathrm{i}}}_{\ell m}(\nu) &= \int_{S^2} \mathrm{d}\Omega(\hat{\mathbfit{n}}) \widetilde{P}^{\,(0)}_{\mathrm{i}}(\hat{\mathbfit{n}}, \alpha_t, \nu) {}_{2}Y_{\ell m}^* (\hat{\mathbfit{n}}) \, , 
\end{split}   \\
\begin{split}
b^{\widetilde{V}^{\,(0)}_{\mathrm{i}}}_{\ell m}(\nu) &= \int_{S^2} \mathrm{d}\Omega(\hat{\mathbfit{n}}) \widetilde{V}^{\,(0)}_{\mathrm{i}}(\hat{\mathbfit{n}}, \alpha_t, \nu) Y_{\ell m}^* \, , %(\hat{\mathbfit{n}}) \, .
\end{split}
\end{align}
\endgroup
to obtain the harmonic coefficients given in Eqs.~\eqref{eq:blm_i}-\eqref{eq:blm_v}.

%%%%%%%%%%%%%%%%%%%%%%%%%%%%%%%%%%%%%%%%%%%%%%%%%%
% Don't change these lines
\bsp	% typesetting comment
\label{lastpage}
\end{document}